%% file: 0_sample-authordraft.tex
\documentclass[sigconf]{acmart}
\AtBeginDocument{%
  \providecommand\BibTeX{{%
    \normalfont B\kern-0.5em{\scshape i\kern-0.25em b}\kern-0.8em\TeX}}}

\copyrightyear{2025}
\acmYear{2025}
\setcopyright{cc}
\setcctype{by}
\acmConference[ASSETS '25]{The 27th International ACM SIGACCESS Conference on Computers and Accessibility}{October 26--29, 2025}{Denver, CO, USA}
\acmBooktitle{The 27th International ACM SIGACCESS Conference on Computers and Accessibility (ASSETS '25), October 26--29, 2025, Denver, CO, USA}
\acmDOI{10.1145/3663547.3746383}
\acmISBN{979-8-4007-0676-9/2025/10}

\usepackage{caption}
\usepackage{subcaption}
\usepackage{longtable}
\usepackage{soul}
\usepackage{multirow}
\usepackage[most]{tcolorbox}
\usepackage{accsupp}

\usepackage{colortbl}
\usepackage{booktabs}
\usepackage{threeparttablex}
\usepackage{enumitem}

\definecolor{pastelorange}{rgb}{1.0, 0.588, 0.3098}

\newcommand{\tick}{\raisebox{-0.2em}{\includegraphics[height=1em]{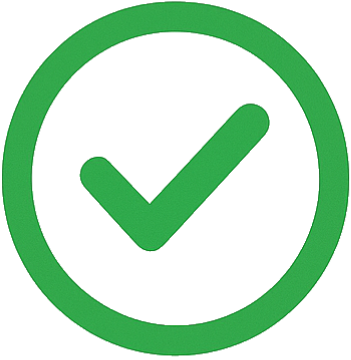}}}
\newcommand{\cross}{\raisebox{-0.2em}{\includegraphics[height=1em]{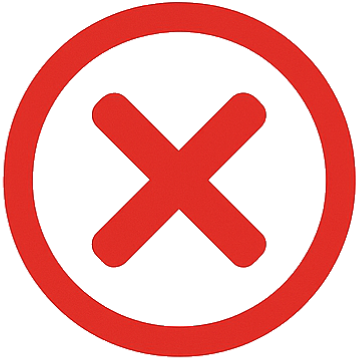}}}

\newcommand{\key}{\raisebox{-0.2em}{\includegraphics[height=1em]{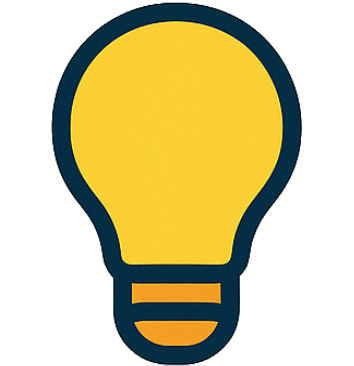}}}

\begin{document}
\title[Smart Glasses for CVI]{Smart Glasses for CVI: Co-Designing Extended Reality Solutions to Support Environmental Perception by People with Cerebral Visual Impairment}


\author{Bhanuka Gamage}
\email{bhanuka.gamage@monash.edu}
\orcid{0000-0003-0502-5883}
\affiliation{%
  \institution{Monash University}
  \streetaddress{Wellington Rd}
  \city{Melbourne}
  \country{Australia}
}

\author{Nicola McDowell}
\email{n.mcdowell@massey.ac.nz}
\orcid{0000-0001-6969-9604}
\affiliation{%
  \institution{Massey University}
  \city{Auckland}
  \country{New Zealand}
}

\author{Dijana Kovacic}
\email{dijana.kovacic@monash.edu}
\orcid{0009-0004-4670-9282}
\affiliation{%
  \institution{Monash University}
  \streetaddress{Wellington Rd}
  \city{Melbourne}
  \country{Australia}
}

\author{Leona Holloway}
\email{leona.holloway@monash.edu}
\orcid{0000-0001-9200-5164}
\affiliation{%
  \institution{Monash University}
  \streetaddress{Wellington Rd}
  \city{Melbourne}
  \country{Australia}
}

\author{Thanh-Toan Do}
\email{toan.do@monash.edu}
\orcid{0000-0002-6249-0848}
\affiliation{%
  \institution{Monash University}
  \streetaddress{Wellington Rd}
  \city{Melbourne}
  \country{Australia}
}

\author{Nicholas Price}
\email{nicholas.price@monash.edu}
\orcid{0000-0001-9404-7704}
\affiliation{%
  \institution{Monash University}
  \streetaddress{Wellington Rd}
  \city{Melbourne}
  \country{Australia}
}

\author{Arthur Lowery}
\email{arthur.lowery@monash.edu}
\orcid{0000-0001-7237-0121}
\affiliation{%
  \institution{Monash University}
  \streetaddress{Wellington Rd}
  \city{Melbourne}
  \country{Australia}
}

\author{Kim Marriott}
\orcid{0000-0002-9813-0377}
\email{kim.marriott@monash.edu}
\affiliation{%
  \institution{Monash University}
  \streetaddress{Wellington Rd}
  \city{Melbourne}
  \country{Australia}
}

\renewcommand{\shortauthors}{Bhanuka Gamage, et al.}

\begin{abstract}
Cerebral Visual Impairment (CVI) is the set to be the leading cause of vision impairment, yet remains underrepresented in assistive technology research. 
Unlike ocular conditions, CVI affects higher-order visual processing—impacting object recognition, facial perception, and attention in complex environments.
This paper presents a co-design study with two adults with CVI investigating how smart glasses, i.e. head-mounted extended reality displays, can support understanding and interaction with the immediate environment. 
Guided by the Double Diamond design framework, we conducted a two-week diary study, two ideation workshops, and ten iterative development sessions using the Apple Vision Pro.
Our findings demonstrate that smart glasses can meaningfully address key challenges in locating objects, reading text, recognising people, engaging in conversations, and managing sensory stress. 
With the rapid advancement of smart glasses and increasing recognition of CVI as a distinct form of vision impairment, this research addresses a timely and under-explored intersection of technology and need.
\end{abstract}



\begin{teaserfigure}
    \centering
  \includegraphics[width=0.87\textwidth,keepaspectratio]{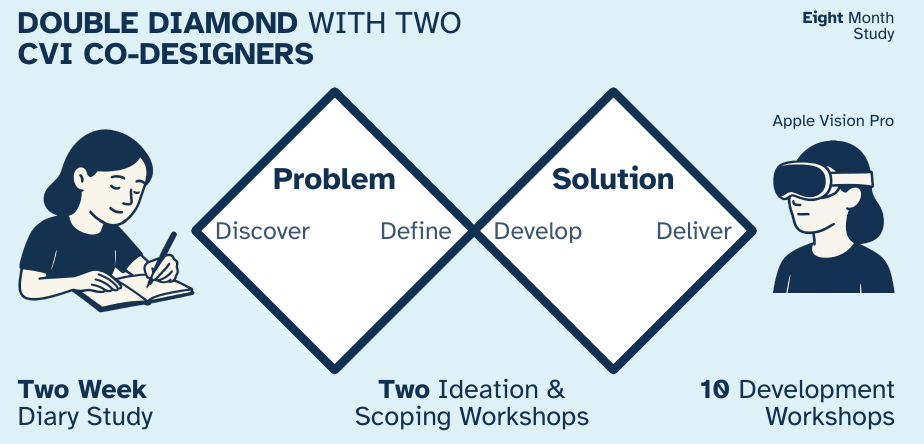}
    \caption{Overview of the study using a Double Diamond co-design process with two co-designers with CVI}
    \Description{Diagram of the Double Diamond with two CVI co-designers, an eight month study. The double diamond consists of two consecutive diamonds 1. Problem. Discover through a two week diary study, then Define. 2. Solution. Develop, then Deliver using an Apple Vision Pro over 10 Development Workshops. The Define and Develop stages involved two ideation \& scoping workshops.}
    \label{fig:full_method_overview}
\end{teaserfigure}


\maketitle


\input{1_introduction}
\input{2_relatedwork}
\input{3_methodology}
\input{4_discover_define}
\input{5_develop_deliver}
\input{6_discussion}
\input{7_limitations}
\input{8_conclusion}

\begin{acks}
We are grateful to the Monash Data Futures Institute for providing the funding that made this project possible.
\end{acks}

\bibliographystyle{ACM-Reference-Format}
\bibliography{sample-base}
\clearpage
\onecolumn
\appendix
\input{9_appendix}
\twocolumn

\end{document}

%% file: 1_introduction.tex
\section{Introduction}

Cerebral Visual Impairment (CVI) is the leading cause of childhood vision impairment in developed countries \cite{sandfeld2007visual, hatton2007babies, matsuba2006long}, and is projected to become a leading cause of adult vision impairment as these children transition into adulthood \cite{bosch2014low}. 
Unlike Ocular Vision Impairment (OVI), which stems from issues with the eye itself, CVI arises from disruptions to the brain’s visual processing centres \cite{sakki2018there}. 
As a result, CVI often affects higher-order visual functions—such as object recognition, face perception, and visual attention—alongside or instead of lower-level functions like acuity or field loss \cite{philip2014identifying, lueck2019profiling}.

Recent advances in extended reality (XR) technologies \cite{kasowski2023systematic} and artificial intelligence \cite{vaswani2017attention, yin2023survey, wu2023multimodal} have opened new possibilities for vision-based assistive tools.
In particular, head-mounted XR displays offer a promising form factor due to their visual, wearable, and hands-free nature \cite{Kim2021Applications}.
In this paper, we use the term “smart glasses” to specifically refer to head-mounted XR displays, such as the Apple Vision Pro \cite{apple_vision_pro}. 
This excludes devices without XR capabilities, such as the Ray-Ban Meta glasses \cite{raybanmeta2024}.
Smart glasses have been widely explored for supporting people with ocular vision impairments \cite{li2022scoping, kasowski2023systematic}, including contrast enhancement, magnification, and navigational overlays \cite{zhao2015foresee, zhao2016cuesee, zhao2020effectiveness}.
While these technologies hold promise, their application to CVI remains largely unexplored.

Emerging work has begun to explore the potential of smart glasses for CVI, identifying key challenges and design opportunities \cite{assets2024, pitt2023strategies, lorenzini2021personalized}.
However these studies have remained at the ideation stage, without advancing to the development or evaluation of real-world solutions.
We therefore ask: \textbf{How can smart glasses alleviate current challenges faced by people with CVI when understanding and interacting with their immediate environment?} 

To address this, we conducted an eight month co-design study with two adults with CVI (who are also co-authors).
Guided by the Double Diamond design framework \cite{banathy2013designing, sanders2008co},  we began with a two-week diary study to uncover their everyday challenges and management strategies.
This was followed by individual ideation and scoping workshops identifying six representative challenges and potential design solutions. 
We then conducted ten iterative development workshops with our co-designers using the Apple Vision Pro to collaboratively prototype, test, and refine these solutions.
Our specific contributions are:
\begin{itemize}
    \item The first co-design study focused on developing and evaluating smart glasses as an assistive platform for adults with CVI.
    \item Evidence that smart glasses can address several key challenges faced by people with CVI when interacting with their environment: locating objects, reading text, recognising people, engaging in conversations, and managing sensory stress. 
    \item Confirmation that visual augmentation of the immediate environment is well suited to people with CVI~\cite{assets2024}, except that there was a preference for challenges involving language to be presented in both text and speech.
\end{itemize}
With the rapid advancement of wearable XR and increasing recognition of CVI as a distinct form of vision impairment, this research addresses a timely and under-explored intersection of technology and need.

%% file: 2_relatedwork.tex
\section{Related Work}

\subsection{Smart Glasses and XR for Visual Assistance}

Smart glasses—also referred to as head-mounted displays (HMDs)— have emerged as a promising form factor within assistive technologies. 
In a systematic review of HMD applications, Kim and Choi \cite{Kim2021Applications} identified 57 studies applying these devices in domains such as surgery, industrial maintenance, and assistive support. 
Their review emphasised the advantages of smart glasses’ visual and hands-free form, particularly in contexts requiring continuous or context-aware interaction.

Focusing on applications for vision impairments, Li et al. \cite{li2022scoping} conducted a scoping review and found 41 studies exploring HMD-based solutions for people with low vision. 
They identified three principal categories of visual enhancement—augmented, modified, and virtual reality—collectively described as extended reality (XR). 
XR refers broadly to immersive technologies that include virtual reality (VR), augmented reality (AR), and mixed reality (MR) \cite{kardong2019call}. 
Li et al. found AR and MR to be especially relevant in assistive contexts where enhancing real-world perception is critical \cite{geringswald2016impairment}, while VR was primarily used for therapeutic or rehabilitative purposes.

Kasowski et al. \cite{kasowski2023systematic} further expanded this landscape by reviewing 76 XR-based studies targeting people with low vision. 
These studies explored a variety of visual augmentation strategies—ranging from contrast enhancement \cite{zhao2015foresee, zhao2016cuesee, zhao2020effectiveness} to spatial cues for navigation \cite{hommaru2020walking, zhao2019designing}—and highlighted the growing potential of XR in augmentations for vision impairments. 
Importantly, Kasowski et al. called for more real-world validation and user-centred design approaches. 
However, their review did not include CVI, despite its growing recognition as a brain-based visual impairment. 
This omission points to a broader gap: while smart glasses have been widely studied for ocular conditions, their application for CVI remains under-explored.


\subsection{CVI-Specific Assistive Technologies}

Despite growing interest in assistive technologies for vision impairments \cite{li2022scoping, lee2021deep, assets2023}, research addressing the specific needs of people with CVI remains limited.
Most existing work focuses on children through case studies \cite{lane2023case, furze2018integrating} or parent-reported experiences \cite{goodenough2021bridging, lupon2018quality}, with minimal exploration of CVI-specific technology use in daily life.
A scoping review by Gamage et al. \cite{gamage2024broadening} found only three studies that directly intersect CVI and assistive technology for daily life.

Early work by Birnbaum et al. \cite{birnbaum2015enhancing} proposed that high-contrast, low spatial frequency visual stimuli could enhance awareness for people with CVI. 
They also speculated on the potential of AR/VR for real-time enhancement, though this did not progress into functional systems.

Pitt and McCarthy \cite{pitt2023strategies} investigated visual scene enhancement for augmentative and alternative communication (AAC) systems. 
While their work did not target CVI specifically, they identified techniques such as colour and light-dark contrast, motion, and scale manipulation as potentially beneficial for this population.

More recently, Lorenzini et al. \cite{lorenzini2021personalized} evaluated the use of portable HMDs in telerehabilitation to reduce device abandonment. 
Although participants with CVI were included, the study primarily focused on ocular impairments and did not explore CVI-specific visual needs or interaction strategies.

Gamage et al.~\cite{assets2024} identified seven key challenges commonly faced by people with CVI: \texttt{C1-Unawareness}, \texttt{C2-Locating}, \allowbreak{} \texttt{C3- Identifying}, \texttt{C4-Reading}, \texttt{C5-Sensory Overload}, \texttt{C6-Mobility} and \texttt{C7-Luminance \& Contrast Sensitivity}. 
Participants suggested various augmentation strategies—such as object highlighting and dynamic filtering—that could potentially support attention and navigation. 
However, the work remained at a conceptual level and emphasised the importance of co-designing solutions in collaboration with people with CVI to ensure validity and adoption.

\subsection{Bridging CVI and XR: Gaps and Opportunities}

Unlike ocular conditions, CVI originates from impairments in the brain rather than damage to the eyes or optic nerve. 
As a result, people with CVI often experience both lower and higher-level visual processing difficulties \cite{philip2014identifying, lueck2019profiling}. 
Gamage et al. \cite{assets2024} emphasised this distinction, outlining five key differences between CVI and low vision: functional vision difficulties, modality preferences, complexity impact on vision, association with other neurological conditions, and the effects of vision rehabilitation. 
They argued that these nuances necessitate tailored approaches and that participatory methods such as co-design are crucial in developing meaningful assistive solutions.

HCI literature has consistently demonstrated the benefits of user involvement in the design of interactive systems, showing improved user satisfaction, better alignment with needs, and faster development cycles \cite{steen2011benefits, kujala2003user}. 
These benefits are particularly critical in the context of assistive technology design \cite{assets2023, li2022scoping}, where personalisation and contextual relevance are often non-negotiable. 
Duckett and Pratt \cite{participant_importance} have also emphasised the importance of participatory research from the perspective of the visually impaired community, who advocate for their active involvement in shaping both the research questions and the design process.

Morrison et al.’s  \cite{Morrison2021} PeopleLens illustrates this approach. 
They developed a head-worn AI system through an extended co-design engagement with a single child who is blind in his daily school environment. 
The researchers were thereby able to iteratively develop features that meaningfully addressed his social and contextual needs—such as identifying classmates and supporting peer interaction. 
Their approach exemplifies the value of deep, context-aware collaboration in developing usable assistive systems.

Although smart glasses are gaining traction in assistive technology, their use for CVI remains largely unexplored \cite{gamage2024broadening, assets2024}. This paper addresses that gap by collaborating with two adults with CVI in an extended co-design process to develop and evaluate smart glasses as assistive tools.
To our knowledge, this is the first study to translate CVI-specific challenges into functioning smart glass solutions through real-world prototyping and iterative collaboration.

%% file: 3_methodology.tex
\begin{table*}[t]
\centering
\renewcommand{\arraystretch}{1.3}
    \begin{threeparttable}
    \caption{Details of the two co-designers.}
    \label{tab:co_designer_details}
        \begin{tabular}{|p{0.8cm}|p{1cm}|p{1cm}|p{0.9cm}|p{1.3cm}|p{6.5cm}|}
        \hline
        & \textbf{Age Group} & \textbf{Gender} & \textbf{Age of \newline Onset} & \textbf{Age of \newline Diagnosis} & \textbf{Visual Difficulties} \newline \textit{(H: Higher-level, L: Lower-level)} \\ \hline
        Dijana & 45 to 54 & Female & Birth & 45 to 54 & \textit{H:} Simultanagnosia\tnote{1}, \newline \textit{L:} Lower Visual Field Impairment \\ \hline
        Nicola & 35 to 44 & Female & 16 & 25 to 34 & \textit{H:} Simultanagnosia, Optic ataxia\tnote{2},\space\space Hemi-attention\tnote{3}\newline \textit{L:} Hemianopsia\tnote{4} \\ \hline
        \end{tabular}
        
        \begin{tablenotes}
        \item[1] Difficulty perceiving multiple visual elements at once.
        \item[2] Impaired reaching for objects using visual guidance.
        \item[3] Reduced attention to one side of space.
        \item[4] Loss of vision in one half of the visual field.
        \end{tablenotes}
    \end{threeparttable}
\end{table*}

\section{Methodology}
This eight-month study followed the Double Diamond design framework \cite{designcouncil2005, banathy2013designing}, a structured, method-agnostic approach to design that emphasises user needs and iterative refinement \cite{Priya2021439, Howard202442, Restianty2024445, Zhang2024}.
This paper is structured around the four phases of the Double Diamond—\textit{Discover}, \textit{Define}, \textit{Develop}, and \textit{Deliver}—which represent alternating divergent and convergent stages of the design process, with each section detailing the methods used.
Within this framework, we adopted a co-design approach~\cite{sanders2008co, morris2024framework} , engaging end-users as equal partners to support shared understanding and ownership of the outcomes.
Figure \ref{fig:full_method_overview} summarises the methodology.

\subsection{Co-designers}
This study involved close collaboration with two adults with CVI. 
We intentionally chose to work with a small group to allow for rich, in-depth and individualised data—prioritising quality and nuance over quantity—which is especially valuable when designing highly personalised and context-aware assistive technologies \cite{Morrison2021, kamath2024playing, schroder2021unboxing}.
This approach reflected that the study was an exploratory investigation aimed at surfacing key design challenges and assessing the feasibility of early prototypes. 

Our first co-designer, \textbf{Dijana}, was formally diagnosed with CVI four years prior to the study, although she has lived with the condition since birth. 
Her contributions reflect the perspective of someone adapting to a diagnosis later in life.
Our second co-designer, \textbf{Nicola}, acquired CVI as a result of a brain injury over twenty years ago. 
Having previously experienced typical vision, Nicola has learned many vision rehabilitation techniques and brings both lived experience and professional expertise to the project as a researcher in the CVI field.

Both had prior collaborations with our team, interest in the technology, and familiarity with design processes. 
They understood that the developed technology might not result in a commercial product—an ethical consideration in AT design \cite{Morrison2021}. 
In recognition of their substantial contributions, both are co-authors of this paper.

The study ran from August 2024 to March 2025. 
Dijana contributed 17 hours, and Nicola 18, across various co-design activities. 
Both were compensated at standard institutional research assistant rates. 
Ethics approval was obtained from our institutional review board. 
Table \ref{tab:co_designer_details} summaries demographic details.

Co-design in this project was a collaborative process in which the two co-designers brought their lived experience, while the research team supported translation into workable technical solutions. While the research team selected the device platform, facilitated workshop sessions, and addressed feasibility constraints, the design decisions were the result of conversations and experimentation between the researchers and co-designers. 
Co-designers identified the most important challenges they faced (Section \ref{sec: diary study}) and added to the initial set of possible solutions during the ideation and scoping workshops (Section \ref{sec: ideation workshop}). 
For instance, the co-designers suggested the user tap on a section of the page to enhance when reading. 
In subsequent workshops (Section \ref{sec: development workshops}), they played a central role in refining prototypes by critically evaluating them and suggesting modifications, such as subtly scaling the text when reading and where to position name tags and arrows. 
This feedback was enhanced by ensuring the co-designers were familiar with the device platform and its capabilities and limitations. 
This was the result of allowing the co-designers to spend considerable time using the device during the ideation workshops and in the subsequent workshops.

\subsection{Device Platform}
\label{sec:device platform}

Selecting an appropriate hardware device and platform was critical to supporting our exploration, particularly given the rapid evolution of head-mounted display (HMD) technology. 
Our primary objective was to investigate specific use cases of high-fidelity vision augmentation for people with CVI. 
As such, we prioritised visual rendering capabilities and system flexibility over immediate practicality or wearability.

Head-mounted visual augmentation systems typically employ either Optical See-Through (OST) or Video See-Through (VST) architectures \cite{langlotz2024design}.
OST systems, such as the Microsoft HoloLens, offer an unmediated view of the physical world but often has low fidelity augmentations. 
In contrast, VST systems—where the user perceives the world through pass-through camera feeds—offer tighter integration of digital content with physical scenes, enabling higher fidelity augmentation at the cost of direct visual access to the environment.

Given the exploratory focus of this study we selected a VST HMD: the Apple Vision Pro \cite{apple_vision_pro_wikipedia}. 
Its ability to deliver high-fidelity, spatially coherent visual overlays was essential for our use cases and justified the trade-off in reduced direct visual clarity associated with VST technology.

The device ran on visionOS 2 and supported development via Apple’s ARKit SDK \cite{apple2025arkit, apple_vision_pro_arkit}, allowing integration of spatial mapping, object tracking, and interactive overlays. 
Additionally, we obtained access to the Vision Pro’s enterprise APIs \cite{apple2024enterpriseapis, apple2024wwdc10139}, granting elevated privileges such as full camera access, custom model execution on the Apple Neural Engine, and increased system resource allocation—capabilities critical for supporting real-time visual processing and experimentation.

Despite its advantages, the platform presented two key limitations:
\begin{itemize}
\item[1.] \textit{Limited eye-tracking data:} 
            Raw eye-tracking data was not accessible; however, the system provided indicators when users gazed at interface elements. 
            This functionality, designed for UI interactions feedback, was not sufficient for detailed gaze analysis but proved useful in some scenarios (as discussed later).
            The lack of fine-grained data limited certain research directions—a constraint shared across most commercial HMDs at the time.
            
\item[2.] \textit{No pass-through manipulation:} The pass-through video feed could not be directly modified—only augmented with overlays. 
    While most commercial VST systems share this limitation, the Varjo XR-4 was a notable exception, offering support for video modification \cite{VarjoVideoPostProcessingAPI}. 
    However, this came at a significant performance cost, often reducing frame rates and introducing latency—trade-offs that could be detrimental for people with CVI.
\end{itemize}

We referred to the device as `glasses' throughout the study to help co-designers envision a future with a less obtrusive form factor, despite the current prototype being a bulkier headset.

%% file: 4_discover_define.tex
\section{Diary Study}
\label{sec: diary study}
We conducted a diary study as part of the \textit{Discover} phase of our double diamond process, aimed at gaining a personal understanding of the two co-designers, their preferences, and the strategies they employ to navigate daily challenges.

\subsection{Procedure}
Diary studies are a well-established method for capturing detailed, time-sensitive, and context-specific data \cite{Carter2005899, Ohly2024151, Jarrahi2021107}. 
Prior work has shown their effectiveness in identifying the practical needs of people with vision impairments, informing the design of more effective assistive technologies \cite{Jeong2020103, Kim201622, AlmgrenBack20241217}.

The diary study spanned two weeks, during which co-designers documented their experiences on four selected days—two weekdays and two weekend days—to capture variation across daily routines.
Entries were recorded using either a physical diary or a mobile app (e.g., Apple Notes). 
Each entry addressed: (1) CVI-related challenges, (2) the activity being performed, (3) the location, and (4) strategies used to manage the challenge.

To support consistency, co-designers were encouraged to set hourly reminders and received daily check-ins from the research team to provide assistance and sustain engagement. 
A sample diary entry provided to the co-designers is shown below:
\begin{tcolorbox}[colback=gray!10, colframe=gray!50, title=\textit{Sample Diary Entry}, fonttitle=\itshape, sharp corners, boxrule=0.5pt]
    \textbf{Date:} **/**/**** \\
    \textbf{Description of the Challenge:} Difficulty reading text on a decorative background on a menu. \\
    \textbf{Activity Being Performed and Location:} Lunch at a local cafe. \\
    \textbf{Management Strategies:} Magnification feature on phone (not successful); asked the server for assistance.
\end{tcolorbox}

Both co-designers intentionally selected four days they felt would yield varied and valuable insights, including activities in unfamiliar settings. 
Dijana included days spent attending an event with her daughter, while Nicola chose days involving out-of-state travel for an academic conference.

\subsection{Data Analysis}
The collected diary entries were compiled and analysed. 
Dijana contributed 18 entries and Nicola 20. 
Three researchers independently categorised each entry by challenge type, which were then grouped into seven high-level categories, identified in \cite{assets2024}.
Any discrepancies in coding were discussed and resolved through consensus.
Subsequently, two researchers conducted a deductive thematic analysis of the strategies co-designers used to manage challenges, identifying four primary strategy types:
\begin{itemize}
\item \textbf{Adapting \& Compromising:} Modifying the approach or accepting a reduced outcome.
\item \textbf{Pre-planning \& Preparation:} Anticipating and preparing in advance.
\item \textbf{Seeking Help \& Asking for Assistance:} Requesting support from others.
\item \textbf{Using Technology \& Tools:} Leveraging devices or tools to complete tasks.
\end{itemize}

\begin{figure*}[!t]
    \centering
    \includegraphics[width=0.95\textwidth]{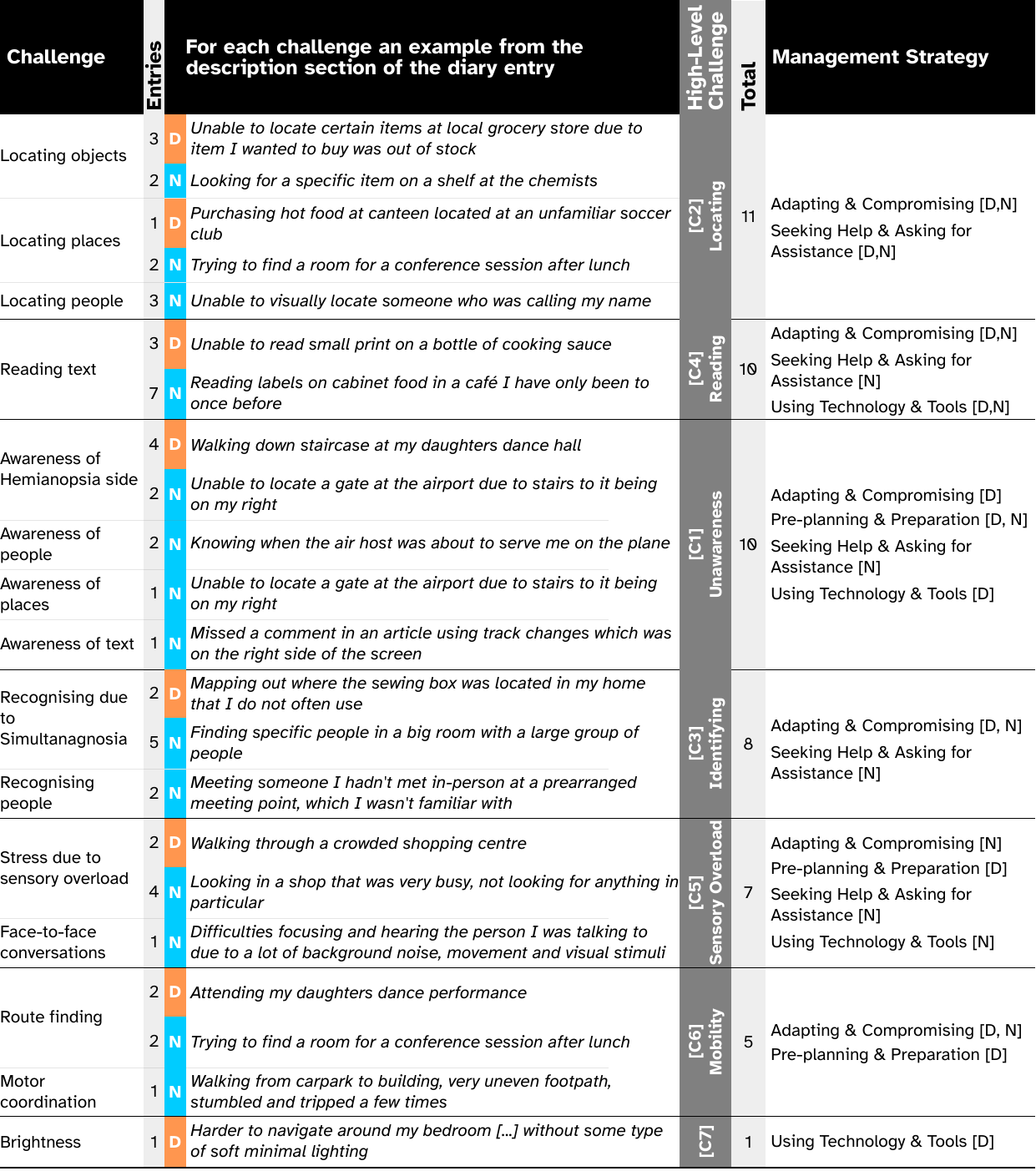}
    \caption{Diary study analysis summarising key challenges experienced by each co-designer, the number of related diary entries, example quotes, corresponding high-level challenges, total occurrence counts, and management strategies. Management strategies are annotated with [D, N] to indicate whether they were used by Dijana, Nicola, or both. [C7] denotes the category of Luminance and Contrast Sensitivity.}
    \Description{Refer to supplementary materials Figure2.csv for the table text.}
    \label{fig:diary_study}
\end{figure*}

\subsection{Challenges}
Our analysis surfaced several recurring challenges that affect everyday functioning for people with CVI.
Figure \ref{fig:diary_study} presents the categorised challenges identified through the diary study.
Some entries were associated with multiple categories, resulting in a total challenge count exceeding the number of entries. 
Notably, all entries mapped to one or more of the seven high-level challenges \cite{assets2024}.

Beyond the nature of individual challenges, the contexts in which they occurred also varied significantly.
Nicola’s entries were often situated in professional environments, including air travel and academic commitments, while Dijana’s challenges were primarily in domestic and caregiving contexts. 
This highlights the diverse situational demands placed on assistive technologies—requiring flexibility across domains of use.

Both co-designers reported feeling stress due to sensory overload in crowded settings, which negatively impacted their concentration and task performance. 
This recurring theme suggests that assistive systems for CVI must not only enhance vision but also reduce sensory overload, a key consideration previous discussed in AR research \cite{duan2022confusing}.

These insights into the challenges faced by each co-designer helped contextualise their everyday experiences and informed our interpretation of how they approached those challenges—laying the groundwork for understanding their management strategies.

\subsection{\textbf{Management Strategies}}

Both co-designers employed personalised strategies to manage CVI-related challenges, though often had to adapt or compromise to complete everyday tasks, as summarised in Figure \ref{fig:diary_study}.
Notably, Dijana tended to minimise reliance on external assistance, instead favouring pre-planning and preparation to help her navigate complex environments. 
She described this approach in the following entry:
\begin{quote}
\textit{Walking through a crowded shopping centre. Pre-planning is required to visit stores where I only need to purchase items and limit my time in the shopping centre to prevent overstimulation.}
\end{quote}

In contrast, Nicola was more inclined to seek help when needed, particularly during unfamiliar situations encountered while travelling during the study. This is reflected in the following entries:
\begin{quote}
\textit{Arrived at destination airport and had to find the taxi stand. I was unable to see the taxi stand sign, so I went back into the airport and asked for directions.}
\end{quote}
\begin{quote}
\textit{Trying to find a room for a conference session after lunch. Gave up trying to locate it myself and followed a colleague.}
\end{quote}

These contrasting approaches—Dijana’s reliance on structured pre-planning and Nicola’s preference for situational assistance—may be shaped by individual personality traits, but are also likely influenced by the time elapsed since the onset of their CVI and the progression of their adaptive strategies. 
Similar patterns have been observed in prior work among people with vision impairments, where people adopt different strategies depending on personal characteristics and their stage of adjustment to vision loss \cite{rai2019coping}.
Importantly, contextual factors also played a role: Nicola was travelling during the study, making advance planning more difficult and reinforcing her need for flexible, situational support.

Understanding these challenges not only revealed the specific barriers faced by each co-designer, but also offered critical insight into the preferences, behaviours, and contexts that shaped their management strategies—fulfilling the diary study’s aim of grounding the design process in the lived experiences of the co-designers.


\section{Ideation and Scoping Workshop}
\label{sec: ideation workshop}

The ideation and scoping workshop addressed two key objectives within the Double Diamond design process.

First, it supported the \textit{Define} phase by helping each co-designer articulate and refine the specific difficulties they face in daily life.
Second, it initiated the \textit{Develop} phase by exploring possible smart glasses–based solutions tailored to those difficulties.

As an outcome of this process, each co-designer selected four difficulties to guide further exploration of possible solutions.
This ensured that the research direction was grounded in their lived experiences, individual challenges, and personal preferences.

\subsection{Procedure}

Each workshop consisted three sessions with each co-designers, facilitated by the primary researcher. 
The session was divided into four parts:

\subsubsection*{\textbf{Discussing Difficulties:}}
The first part focused on defining the specific difficulties co-designers face in their daily lives. 
The session began by revisiting the challenges documented in the diary study.
Co-designers reviewed, modified, and extended this list with additional difficulties drawn from their own lived experience. 
For completeness, we then incorporated a set of difficulties identified in prior work \cite{assets2024}, which was based on a larger cohort of people with CVI.
The finalised list is presented in the “Difficulties” column of Figure \ref{fig:difficulties_possible_solutions} (in the Appendix).

\subsubsection*{\textbf{Introduction to the Device Platform:}}

Following the definition of difficulties—marking the transition from the \textit{Define} to \textit{Develop} phase—co-designers were introduced to the Apple Vision Pro. 
As this was their first time using the device, the researcher guided them through its on-boarding and calibration process. 
Co-designers then explored several demonstration applications to explore the device’s augmented reality capabilities.
Afterwards, they shared initial impressions and feedback.

\subsubsection*{\textbf{Ideating Possible Solutions:}}

The third part focused on generating possible smart glasses–based solutions for the defined difficulties. 
A possible solution was defined as a viable design concept that could address a specific difficulty using the capabilities of the device. 
For example, for the difficulty “Locating items in a cluttered environment,” possible solutions included “Highlighting the object” or “Providing auditory directions.”
The ideation process was grounded in the co-designers’ direct, hands-on experience with the device.
To expand the design space, we also explored opportunity areas identified in prior work \cite{assets2024}.
The resulting possible solutions are presented in the “Possible Solutions” column of Figure~\ref{fig:difficulties_possible_solutions} (in the Appendix).

\subsubsection*{\textbf{Selecting Difficulties to Explore Further:}}

To identify which solutions were of greatest potential benefit to our co-designers, they rated the usefulness of all previously ideated solutions on a 3-point scale (1=not useful, 2=somewhat useful, 3=very useful). 
We then filtered the original list of difficulties—identified during the workshop phase—to include only those that had at least one associated solution rated as `very useful'. 
This process prioritised difficulties that were both relevant to the co-designers and seen as addressable by smart glasses.

Following this, each co-designer selected four key difficulties to carry forward into the development phase, in collaboration with the primary researcher.
These selections were primarily guided by the co-designers’ lived experiences, priorities, and motivation to continue exploring the challenges. 
The researcher provided additional input regarding technical feasibility and novelty, ensuring that the selected directions were both actionable within the device’s capabilities and under-explored in existing literature.

For example, difficulties related to `Brightness' (\texttt{C7–Luminance \& Contrast Sensitivity}) were excluded due to hardware constraints that prevented adjustments to screen brightness and pass-through video. 
Similarly, `Route finding' (\texttt{C6–Mobility}) was excluded, as this is already well-addressed by academic \cite{zhao2020effectiveness, qiu2023use} and commercial solutions \cite{google_maps_live_view, hyperar2025}.

By making sure the final selections were personally meaningful to the co-designers and practically viable, this process laid a strong foundation for the development phase that followed.

\subsection{Identified Difficulties}
During the workshop discussions, co-designer's revealed two difficulties that were not identified in either the diary study or prior work \cite{assets2024}. 
This underscores the value of the workshop as a complementary method for surfacing more nuanced, context-dependent challenges.

The first new difficulty related to processing auditory and visual information simultaneously, as described by Nicola:
\begin{quote}
\textit{"If you want someone with CVI to see you, don’t expect them to hear you, and if you want them to hear you, don’t expect them to see you."}
\end{quote}
This reflects a sensory integration issue, in which the cognitive demands of managing multiple modalities can impair visual processing. 
While this phenomenon is commonly observed in lived experience with CVI, it remains underrepresented in clinical assessments and standard low vision evaluation protocols \cite{kran2019cerebral}. 
Following discussion, this difficulty was categorised under \texttt{C5–Sensory Overload}.

The second difficulty, raised by Dijana, involved sensitivity to blue light, particularly flickering light sources such as screens and emergency vehicle sirens. 
While specific sensitivity to blue light is less documented for people with CVI, Photophobia\footnote{Difficulty tolerating light} has been reported in approximately one-third of children with CVI \cite{jan1993photophobia}. 
Recent studies have also explored the use of assistive technology to mitigate light sensitivity in broader populations \cite{ghosh2023artificial, hu2022dc, hu2024smart}. 
After discussion, this challenge was categorised under \texttt{C7–Luminance \& Contrast Sensitivity}.

The full, updated list of 20 difficulties considered during the workshop is presented in the Difficulties column of Figure~\ref{fig:difficulties_possible_solutions} (in the Appendix).

\subsection{Feedback on the Device Platform}

Both co-designers expressed enthusiasm for the Apple Vision Pro, particularly its immersive augmented reality features, and recognised its potential as a visual assistance tool. 
At the same time, both noted the device’s bulky form factor as a barrier to everyday use. 
Despite this limitation, they were optimistic about future iterations, anticipating that ongoing advancements in wearable technology would lead to more compact, socially acceptable designs. 
This insight guided discussions toward designing for a future scenario where such capabilities may exist in everyday glasses-like devices.

\subsection{Ideated Possible Solutions}
During this stage, both co-designers contributed actively to the ideation process, generating a range of possible solutions tailored to the previously defined difficulties. 
Many of these ideas were grounded in their lived experiences and informed by their first-hand interaction with the device platform.

For instance, Nicola proposed enhancing text readability by reformatting printed text one word at a time, triggered by eye-gaze—an approach aimed at reducing visual clutter and improving focus. 
She also suggested using the glasses to assist with attention management by selectively muting or dimming irrelevant auditory and visual stimuli to help focus on a specific person or object.
Dijana offered a solution to mitigate her sensitivity to blue light, proposing that the glasses digitally transform intense blue hues—such as those from screens or sirens—into darker, less problematic tones.

In total, 45 distinct possible solutions were generated across the 20 identified difficulties (as detailed in the Possible Solutions column of Figure~\ref{fig:difficulties_possible_solutions} in the Appendix).

\subsection{Selected Difficulties for Development}
\begin{figure}[htbp]
    \centering
    \includegraphics[width=\linewidth]{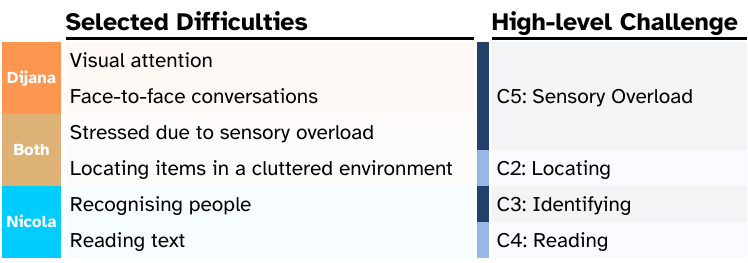}
    \caption{Selected difficulties identified by each co-designer to guide the focus of smart-glass solution development during the development workshops. Difficulties are colour-coded by co-designer (Dijana: pastel orange, Nicola: cyan, Both: brown) and mapped to corresponding high-level challenges.}
    \label{fig:selected_difficulties}
    \Description{Table with the headings Co-designer: Selected Difficulties; High Level Challenge.
        Dijana: Visual Attention; C5 Sensory Overload
        Dijana: Face to face conversations; C5 Sensory Overload
        Both: Stressed due to sensory overload; C5 Sensory Overload
        Both: Locating items in a cluttered environment; C2 Locating
        Nicola: Recognising people; C3 Identifying
        Nicola: Reading text; C4 Reading
        }
\end{figure}

Through collaborative discussion sessions, each co-designer selected four key difficulties to carry forward into the development phase.
Notably, two difficulties were selected by both co-designers, while the remaining two reflected each person’s unique lived experience.
As shown in Figure~\ref{fig:selected_difficulties}, these six final difficulties span four of the seven high-level challenges.
These six difficulties shaped the direction of the subsequent \textit{Develop} and \textit{Deliver} phases, guiding the ten co-design development workshops described in the following sections.
Beyond defining the problem space, the ideation and scoping workshop established a collaborative design partnership—centreing the co-designers’ lived experiences and priorities to ensure that development efforts remained both meaningful and actionable.

%% file: 5_develop_deliver.tex
\begin{figure*}[t]
    \centering
    \includegraphics[width=\textwidth]{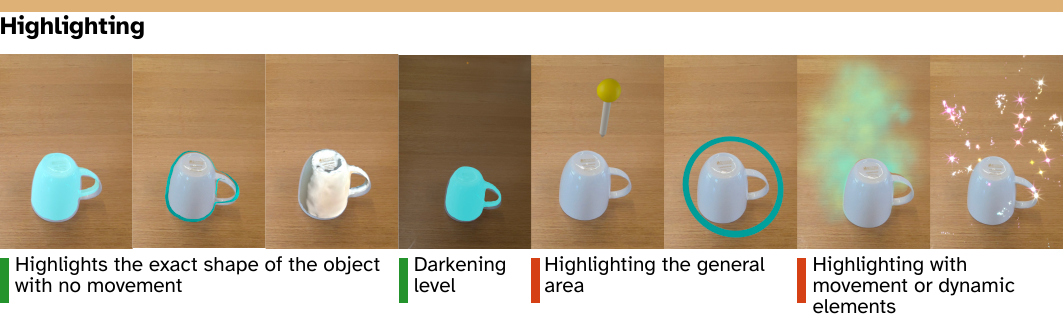}
    \caption{Examples of the 28 design options explored to address the difficulty of locating items in a cluttered environment. 
    Green bars indicate options that were liked by both co-designers, while red bars indicate options that were not preferred. }
    \label{fig:possible_solutions_highlighting}
    \Description{A series of images relating to Highlighting, each with a caption and a red or green bar to indicate whether it was liked by both co-designers. The description gives the caption, (colour), image description.
            Highlighting
            Highlights the exact shape of the object with no movement (green). Images of a mug with cyan highlight over the whole mug, outlining the mug, or replacing the mug with a 3D rendering.
            Darkening level (green). Image of a mug with cyan highlight over the whole mug and a darkened background for increased contrast.
            Highlighting in the general area (red). Images of a mug with a pin hovering above it, or with a cyan circle around it.
            Highlighting with movements or dynamic elements (red). Image of mug with sparkles shooting outwards.
            }
\end{figure*}

\section{Development Workshops}
\label{sec: development workshops}

The development workshops aimed to collaboratively design and refine solutions for the six key difficulties identified by the co-designers. 
Figure \ref{fig:methodology_overview} (in the Appendix) shows the primary focus of each workshop. 
While earlier ideas were revisited and refined in later sessions, the figure highlights the main design explorations per workshop.

Two of the four selected difficulties were specific to each co-designer, so some solutions were initially developed with only one participant. 
To ensure broader feedback, these were later shared with the other co-designer. 
For instance, solutions for `Reading Text’—developed with Nicola—were presented to Dijana during Workshop V for feedback.

\subsection{Procedure}

The development workshops took place at the co-designers’ homes and were primarily led by the first author, supported by another researcher from the team. 
Each session lasted between two to four hours. 
Co-designers were encouraged to take breaks as needed and actively discuss how well the possible solutions worked, identifying strengths and areas needing improvement.
Each workshop typically followed these three stages:

\subsubsection*{\textbf{Discussion of Possible Solution:}}
At the beginning of each session, the first author presented the solution to the co-designers. 
This was done with an active discussion, supported by a set of guiding questions designed to prompt deeper engagement—for example, whether any visual elements should be modified or if the motion and behaviour of specific overlays felt intuitive. 
Feedback and discussions were recorded while the co-designers interacted with the system.

\subsubsection*{\textbf{Modifications Based on Feedback:}}
The team when possible, made immediate changes to the solution. 
For instance, if a visual component appeared too small or if co-designers suggested a different colour, adjustments were implemented immediately. 
This iterative approach enabled rapid improvements and immediate evaluation of modifications.

\subsubsection*{\textbf{Evaluation of User Experience:}}
Conducting workshops in the co-designers’ homes allowed assessments in a familiar and realistic environment. 
To evaluate how effectively the solutions worked in practice, specific scenarios reflecting actual usage were recreated. 
Furthermore, discussions also considered how the solutions might function in less familiar contexts, such as public settings.

\subsection{Design Exploration}

This section details the design exploration phase, focusing on the possible solutions for the six previously identified difficulties. 
Each subsection summarises what the co-designers liked \tick\space and disliked \cross, and the key takeaway \key.


\subsubsection*{\textbf{Locating items in a cluttered environment:}}
\label{sec:design_exploration:locating}
A total of 28 design options were explored across two areas: informing users about an object already within their view, and directing the user towards the object if it was out of view. 
Techniques explored were inspired by prior design studies in assistive technology \cite{zhao2016cuesee, cookar} and HCI \cite{gruenefeld2019locating, dogan2024augmented, zhang2022initial}. 

\begin{itemize}[label=\tick, leftmargin=*, itemsep=2pt]
    \item Co-designers preferred visual methods that directly highlighted or outlined the target object using their preferred colours--Dijana selected pastel orange, while the Nicola preferred cyan.
    Figure \ref{fig:possible_solutions_highlighting} illustrates examples of these techniques. 
    Nicola highlighted the additional benefit for users with optic ataxia:
        \begin{quote}
        \textit{The highlighting not only helps draw visual attention, but it also helps with optic ataxia. It's because its actually outlining the exact shape more clearly so I know the exact size of the object to grab.}
        \end{quote}
    
    \item Darkening the surrounding environment to highlight the target object was also positively received: 
    \begin{quote}
        \textit{The clutter being darkened out and makes the object very obvious. Interestingly the darkening has a calming effect.}
        \end{quote}
        \begin{quote}
    \end{quote}
    
    \item For objects outside the immediate field of view, co-designers preferred a large arrow pointing toward the target, offset from their central vision. Dijana described the benefit clearly:
        \begin{quote}
            \textit{This approach works better because the arrow is offset and peripheral. It should disappear once facing the object.}
        \end{quote}
\end{itemize}

\begin{itemize}[label=\cross, leftmargin=*, itemsep=2pt]
    \item Co-designers disliked visuals pointing generally to areas rather than directly to objects, especially those involving movement or constantly changing indicators. Nicola explained this clearly:
    \begin{quote}
    \textit{"When I see that pointer, it might not be where it actually is, you will need to then figure out where it's pointing to."}
    \end{quote}
    
\item Glowing indicators at the edges of vision as a way to direct the user towards an object was found to be irritating and ineffective.

\item Auditory methods, either alone or combined with visuals, were not preferred. These included spatial sounds from the object's location and spoken directions telling the user where to turn. 
 Nicola explained the difficulty in figuring out where sounds were coming from, especially in noisy places:
\begin{quote}
\textit{"I can't visually locate where the sound is coming from. I don't think it will be very reliable. I would just be looking around trying to figure out where the sound is coming from."}
\end{quote}
\end{itemize}

\begin{itemize}[label=\key, leftmargin=*, itemsep=2pt]
\item Co-designers found the smart glasses were effective for locating items in cluttered environments. 
    They preferred stable visual highlights over audio, found darkening improved focus and comfort, and favoured peripheral arrows to guide attention toward objects outside the field of view.
\end{itemize}

\begin{figure*}[t]
    \centering
    \includegraphics[width=\textwidth]{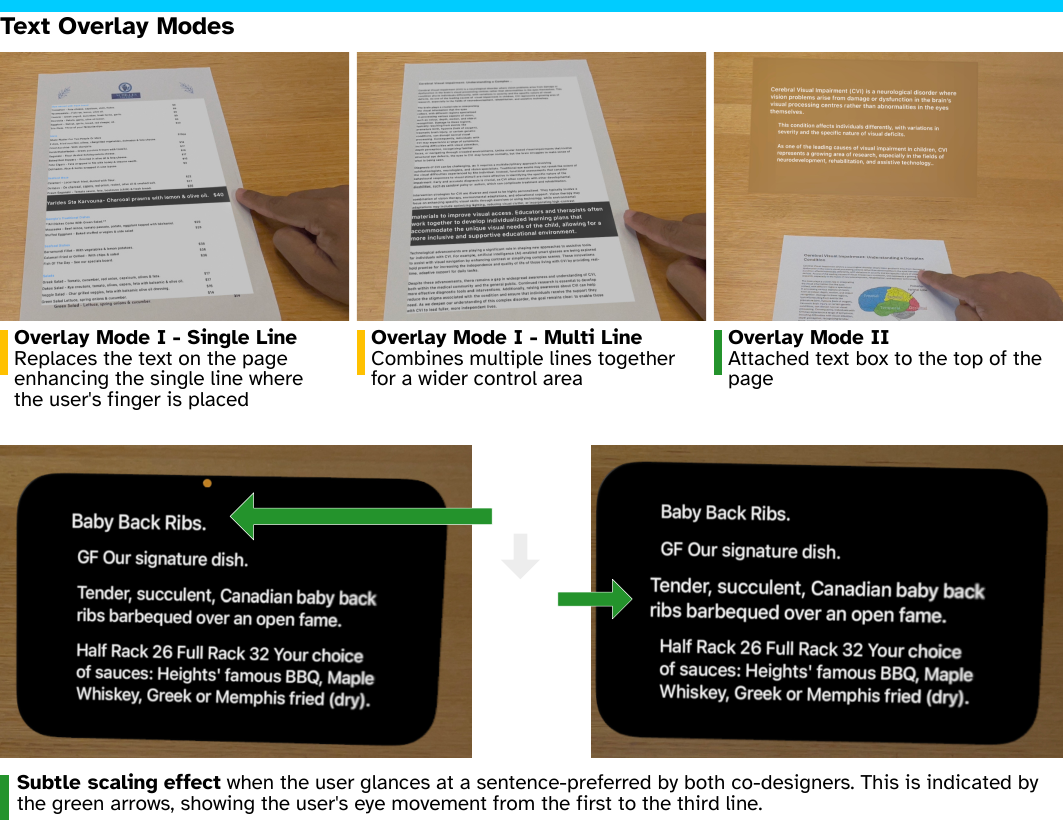}
    \caption{Examples of the 13 design options explored to support the difficulty of reading text. The top row presents different text overlay modes, while the bottom row shows a subtle scaling effect triggered by eye gaze movement. Green bars indicate options the user liked, whereas yellow bars indicate options that were effective but ultimately not preferred due to practical limitations.}
    \label{fig:possible_solutions_reading_text}
    \Description{Images illustrating text overlay modes, each with a caption and a green bar to indicate that it was liked by both co-designers or a yellow bar to indicate that it was ultimately not preferred. The description gives the caption, (colour), image description.
    Overlay Mode I - Single Line. Replaces the text on the page enhancing the single line where the user’s finger is placed (yellow). Photograph of a page with text replaced on top, with clear black print on a white background. A finger points to a single line, which is highlighted as white text on a black background.
    Overlay Mode I - Multi Line. Combines multiple lines together for a wider control (yellow). Similar image, but with a whole paragraph of text highlighted.
    Overlay Mode II. Attached text box to the top of the page (green). A finger points to a print page of text. Above the page, there is a floating box with clear print.
    Subtle scaling effect when the user glances at a sentence - preferred by both co-designers. This is indicated by the green arrows, showing the user’s eye movement from the first to the third line (green).  Close up images of an overlay text box. In the first image, an arrow points to the top line, which is in slightly larger text. In the second image, an arrow points to the third paragraph, which is now the text in larger font.  
    }
\end{figure*}
\begin{figure*}[t]
    \centering
    \includegraphics[width=\textwidth]{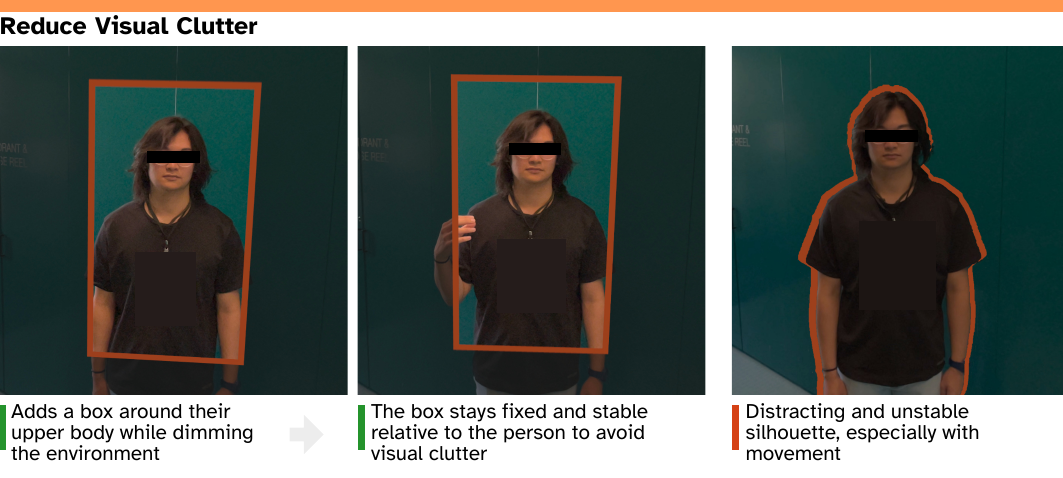}
    \caption{Examples of design options explored to address the difficulty of face-to-face conversations. Green bars indicate options that were liked by both co-designers, while red bars indicate options that were not preferred. }
    \label{fig:possible_solutions_face_to_face}
    \Description{Images illustrating “Reduce Visual Clutter”, each with a caption and a coloured bar to indicate that it was liked by both co-designers (green) or not preferred (red). The description gives the caption, (colour), image description.
    Adds a box around their upper body while dimming the environment (green). A person facing the viewer with an orange rectangle around their head and torso. The area outside the rectangle is dimmed.
    The box stays fixed and stable relative to the person to avoid visual clutter (green). The person has raised their arm but the box remains in the same position.
    Distracting and unstable silhouette, especially with movement (red). An orange line outlines the person’s head and body. All background is dimmed.
    }
\end{figure*}
\subsubsection*{\textbf{Reading text:}}
We explored 13 design options across two categories: visual augmentations and text-to-speech. 
Techniques explored were inspired by prior design studies in assistive technology, mainly for people with low-vision \cite{zhao2015foresee}.

\begin{itemize}[label=\tick, leftmargin=*, itemsep=2pt]
    \item Both co-designers reacted positively to the text-to-speech solution, which enables users to tap any section of text to hear it read aloud. 
    Nicola described this feature as potentially:
        \begin{quote}
        \textit{"Life changing, this is awesome, I would use this a lot. I have stopped reading magazines, but I would go back to reading magazines because of this."}
        \end{quote}

    \item Co-designers strongly preferred a combined solution that integrated text-to-speech with Overlay Mode II—an interface that presents enhanced text at the top of the page (see Figure \ref{fig:possible_solutions_reading_text}). 
    They found this multimodal design particularly effective in noisy environments such as restaurants or supermarkets, where they relied primarily on auditory information.
    Importantly, the visual overlay was not used as the primary channel for consuming content in these settings. 
    Instead, it served as a form of system feedback—confirming what the device had recognised and was reading aloud. 
    This distinction highlights a critical nuance in modality design: although the system offered both visual and auditory outputs, co-designers used them in a complementary, sequential way—not simultaneously for parallel information processing.

    
    \item An interesting observation arose during the exploration of Overlay Mode II (see Figure \ref{fig:possible_solutions_reading_text}). 
    A feature termed `Subtle Scaling', which uses the device platform's visual feedback system (described under limitations in Section \ref{sec:device platform}) was added.
    This feature slightly enlarges the text line the user is viewing (illustrated in Figure \ref{fig:possible_solutions_reading_text}), which guided visual attention without overwhelming the field of view.
    Both co-designers reported that this subtle scaling was effective and found it improved their reading experience.
    
    \item Both co-designers valued the customisability of visual settings, such as font type, size, colour, background colour, and line spacing, as well as text-to-speech options including playback speed and voice selection. 
    These preferences align with findings from prior research for children with CVI \cite{smolansky2024towards}.

\end{itemize}

\begin{itemize}[label=\cross, leftmargin=*, itemsep=2pt]
\item While both co-designers found Overlay Mode I effective, they struggled with hand-tracking precision due to optic ataxia (see Overlay Mode I – Single Line in Figure \ref{fig:possible_solutions_reading_text}). 
    To mitigate this, we expanded the interactive control area to reduce the need for fine motor input (see Overlay Mode I – Multi Line in Figure \ref{fig:possible_solutions_reading_text}). 
    However, even with this adjustment, the interaction remained challenging. 
    Both co-designers suggested that smooth eye-tracking would provide a more accessible alternative, but this could not be implemented due to current hardware limitations.
\end{itemize}

\begin{itemize}[label=\key, leftmargin=*, itemsep=2pt]
\item Smart glasses was identified as a powerful tool for reading through visual overlays and customisable text-to-speech. 
Preferences varied by context, underscoring the importance of adaptive, multimodal support in everyday environments.
\end{itemize}

\subsubsection*{\textbf{Face-to-face conversations:}}
We explored six design options across two strategies—reducing visual clutter and background noise—to support focus during face-to-face conversations.

\begin{itemize}[label=\tick, leftmargin=*, itemsep=2pt]
    \item Both co-designers reported that drawing a virtual rectangular frame around the upper body of the conversation partner, coupled with dimming the surrounding environment, improved their ability to focus on the speaker (see Figure \ref{fig:possible_solutions_face_to_face}). 
    This approach was especially effective for Dijana, who described a marked improvement in her visual integration:
    \begin{quote}
    \textit{"Something in my visual pathway falls into place because of this. You were able to recreate how it feels if my CVI went away. I was able to start the process of integrating those senses."}
    \end{quote}
    Further exploration of this design revealed that the reduction in visual clutter and the presence of a defined outline around the speaker helped direct attention more effectively. 
    As Dijana noted, \textit{“Dark lighting helps reduce the visual clutter, the outline helps me focus on the person.”}

    \item Both co-designers expressed a preference for a stable, fixed rectangular frame rather than a dynamically adjusting cutout that precisely traced the speaker’s silhouette. 
    The latter was found to be visually distracting due to continuous updates triggered by small movements. 
    In contrast, a static box allowed for natural body motion without requiring constant visual adjustments.

    \item In scenarios involving multiple speakers, Dijana initially proposed combining individual frames into a single larger one. 
    However, after further discussion, both co-designers favoured an adaptive approach where the system highlights only the current speaker. 
    This dynamic switching was perceived as more effective for tracking conversations.

    \item In terms of reducing background noise, the co-designers differed in their preferences. 
    Nicola favoured the use of active noise cancellation, while Dijana preferred an unaltered auditory environment.
    To evaluate these preferences, Apple AirPods Pro \cite{wiki:airpods} were connected with the device platform to enable selective noise reduction. 
    Nicola observed that for her:
    \begin{quote}
    \textit{"Adding the noise cancellation has a bigger effect for the conversation compared to the box around the person."}
    \end{quote}
    These differing preferences suggest that future systems should offer users flexible control over audio settings, enabling them to adapt the experience to their specific sensory needs and environments.

    \item During the sessions, Nicola proposed an additional idea: displaying real-time captions during speech to support comprehension. 
    There is substantial existing research on real-time captioning and translation for hearing accessibility \cite{peng2018speechbubbles, jain2018towards, jain2018exploring}, and Li et al. \cite{li2023augmented} have also contributed a detailed design space analysis in this area. 
    However, due to time constraints, we did not explore this direction further in our study.    
\end{itemize}

\begin{figure*}[t]
    \centering
    \includegraphics[width=\textwidth]{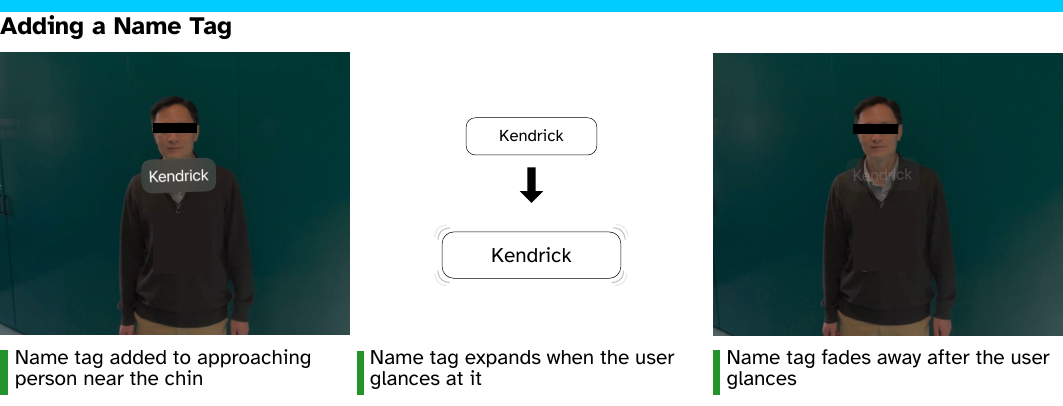}
    \caption{Examples of design options explored to support the difficulty of recognising people. Green bars indicate options that were liked by both co-designers. }
    \label{fig:possible_solutions_recognise_people}
    \Description{Images illustrating “Adding a name Tag”, each with a caption and a green bar to indicate that it was liked. The description gives the caption, (colour), image description.
    Name tag added to approaching person near the chin (green). Person with a name tag floating beneath their chin.
    Name tag expands when the user glances at it (green). Image of name tag increasing in size.
    Name tag fades away after the user glances (green). Person with name tag removed.
    }
\end{figure*}

\begin{itemize}[label=\cross, leftmargin=*, itemsep=2pt]
    \item Despite the effectiveness of reducing visual clutter, Nicola noted that she had independently developed personal cognitive strategies over time to achieve similar outcomes:
    \begin{quote}
    \textit{"If you have given this to me [many] years ago, this would have been awesome. Essentially what you have done, it's brilliant, but I've done that myself mentally through other strategies. It will be very useful for someone who has newly acquired CVI. I know which things distract me and has trained my brain."}
    \end{quote}
    This insight suggests that such interventions may be more valuable for people with newly acquired CVI, whereas those with longstanding experience may have developed management strategies.
\end{itemize}

\begin{itemize}[label=\key, leftmargin=*, itemsep=2pt]
\item Smart glasses supported face-to-face conversations by reducing visual clutter. 
Co-designers preferred dimmed surroundings and a stable virtual box around the speaker, which helped sustain focus—particularly useful for people with newly acquired CVI in social settings.
\end{itemize}

\begin{figure*}[t]
    \centering
    \includegraphics[width=\textwidth]{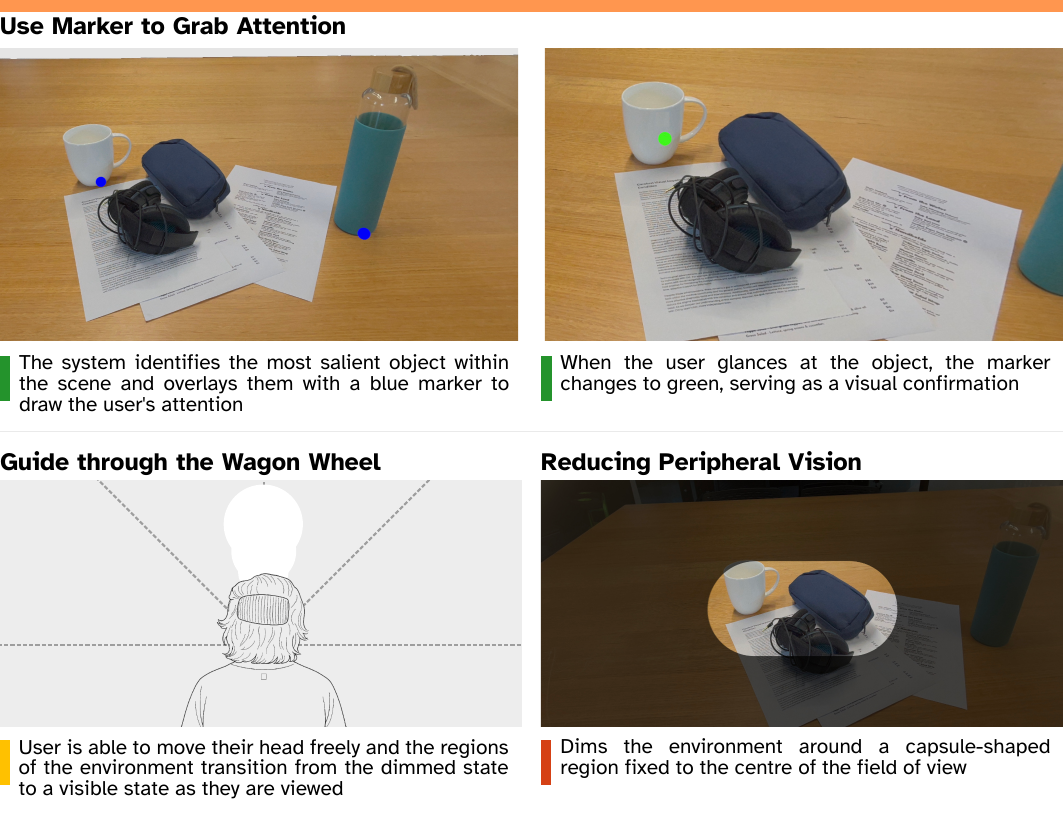}
    \caption{Examples of the 14 design options explored to support the difficulty of Visual Attention. Illustrated options include virtual markers for guiding attention, dynamic dimming to reduce distractions, and spatial techniques like the `wagon wheel' to progressively reveal visual information based on head orientation. Green bars indicate options that were liked by both co-designers, while red bars indicate options that were not preferred. Yellow bars indicate options that were effective but ultimately not preferred.}
    \label{fig:possible_solutions_visual_attention}
    \Description{Images illustrating “Use marker to grab attention”, “guide through the wagon wheel” and “reducing peripheral vision, each with a caption and a coloured bar. The description gives the caption, (colour), image description.
    Use Marker to Grab Attention
    The system identifies the most salient object within the scene and overlays them with a blue marker to draw the user’s attention (green). Items on a desk.
    When the user glances at the object, the marker changes to green, serving as a visual confirmation (green). Small dots are placed on or near a mug (blue), case (orange) and water bottle (blue).
    When the user glances at the object, the marker turns to green, serving as a visual confirmation (green). The mug now has an orange and a green dot overlaid on top of it.
    Guide through the Wagon Wheel
    User is able to move their head freely and the regions of the environment transition from the dimmed state to a visible state as they are viewed (yellow). Diagram of a person wearing a VR headset. The environment is dimmed out except for an area above their head, which they have already viewed.
    Reducing Peripheral Vision
    Dims the environment around a capsule-shaped region fixed to the centre of the field of view (red). Items on a desk with a capsule-shaped window around the items and dimming outside the capsule.
    }
\end{figure*}

\subsubsection*{\textbf{Recognising People:}}
Two design options were developed to support person recognition using both visual and auditory modalities.
Techniques explored were inspired by prior work in XR that explores name tag placement on both objects \cite{azuma2003evaluating} and people \cite{wang2023flying, stevens2018social}.

\begin{itemize}[label=\tick, leftmargin=*, itemsep=2pt]
    \item Both co-designers responded positively to the idea of displaying virtual name tags for people in the environment, rather than relying solely on auditory identification cues. 
    In the initial design, the name tag was positioned above the person’s head. However, Nicola raised concerns about the practicality and accessibility of this placement:
    \begin{quote}
    \textit{"When I meet someone, I don't want to look up, it's a bit unnatural. This would also affect people with visual field issues [hemianopsia], if its too high above the person's head."}
    \end{quote}
    In response, the tag’s position was revised to appear just below the chin, enabling easier visibility through a brief downward glance (see Figure \ref{fig:possible_solutions_recognise_people}).
    
    \item The co-designers emphasised the importance of the system being able to automatically recognise familiar people and display their name tags without requiring any explicit user input. 
    Nicola illustrated a common scenario where this functionality would be particularly beneficial:
    \begin{quote}
    \textit{"The biggest [use case] for me is when people are walking towards me, and they are coming to talk to me, and then I am like who is this?"}
    \end{quote}

    \item Both co-designers emphasised the importance of name tags that adapt dynamically based on user interaction and context. 
    To address this, we prototyped behaviours such as \textit{Scaling on Glance}, where the name tag briefly enlarges upon direct gaze, and \textit{Fade After Recognition}, where it gradually disappears after being viewed. 
    These adaptations aim to provide timely identity cues while minimising visual clutter, aligning with cognitive load reduction principles in assistive interface design \cite{lang2021pressing, kalatzis2023multimodal}.

\end{itemize}

\begin{itemize}[label=\cross, leftmargin=*, itemsep=2pt]
    \item Despite the perceived usefulness of virtual name tags, both co-designers expressed concerns about potential information overload. 
    Specifically, they did not want the device to continuously display name tags for every detected person, which could lead to unnecessary visual clutter. 
    To address this, the idea of \textit{Photo-Based Recognition} was proposed, allowing users to contribute photos of known people to help train the glasses. 
    This approach could improve recognition accuracy while allowing the user greater control over which people are identified in the interface.
\end{itemize}


\begin{itemize}[label=\key, leftmargin=*, itemsep=2pt]
\item Both co-designers found the virtual name tag to be a subtle yet effective tool for recognising people.
\end{itemize}

\subsubsection*{\textbf{Visual attention:}}
We explored 14 design options across five solution areas aimed at guiding visual attention.
Design options explored were inspired by other studies in the field, which have used similar methods for guiding visual search \cite{lu2012subtle} and navigation \cite{koch2014natural}. 

\begin{itemize}[label=\tick, leftmargin=*, itemsep=2pt]

    \item One proposed solution involved adding visual markers to highlight the most salient items within the user’s field of view (see Figure \ref{fig:possible_solutions_visual_attention}). 
    When a user glanced at a marked item, the colour of the marker would change to indicate that the object had been visually seen. 
    Both co-designers responded positively to this feature. 
    Dijana remarked that the markers were helpful for drawing attention: \textit{“It’s good, it’s more attention grabbing.”}
    While both co-designers found visual markers initially helpful for attracting attention, they cautioned that excessive or persistent prompts could become distracting or fade into the background.

    \item Another solution was inspired by the ‘wagon wheel’ strategy, a technique used by some people with CVI to incrementally build a visual understanding of a scene \cite{mcdowell2021wagon}. 
    This method involves moving the gaze outward from a central fixation point in a radial pattern—similar to the spokes of a wagon wheel—while consistently returning to the centre. 
    To support this, a visual aid was designed where the glasses would initially dim the full scene, then progressively reveal circular zones expanding from the centre. 
    This gave users a visual map of which areas had already been scanned (see Figure \ref{fig:possible_solutions_visual_attention}). 
    Nicola strongly supported this approach, noting its potential to help users learn how to systematically build a mental map of their surroundings.
\end{itemize}

\begin{figure*}[t]
    \centering
    \includegraphics[width=\textwidth]{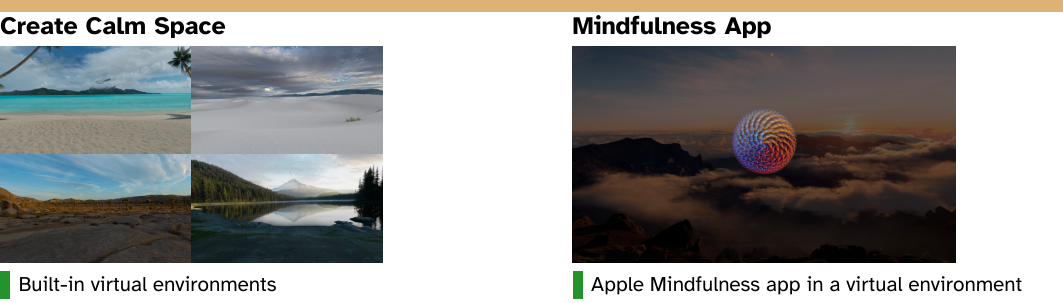}
    \caption{Examples of design options explored to address stress caused by sensory overload. 
    Strategies include (left) presenting calming virtual environments, and (right) integrating the Apple Mindfulness app within an immersive setting to support emotional regulation. Green bars indicate options that were liked by both co-designers.}
    \label{fig:possible_solutions_stressed}
    \Description{Images illustrating “Create Calm Space” and “Mindfulness App” , each with a caption and a coloured bar. The description gives the caption, (colour), image description.
    Create Calm Space
    Built-in virtual environments (green). Four high-quality images of landscapes: beach with water and palm trees, sand dunes, rocky mountains, and a lake with trees and mountains in the background.  
    Mindfulness App
    Apple Mindfulness app in a virtual environment (green) multicoloured disk floating above clouds at sunset.
    }
\end{figure*}

\begin{itemize}[label=\cross, leftmargin=*, itemsep=2pt]
    \item Both co-designers strongly disliked the approaches that direct visual attention deliberately such as reducing peripheral vision (see Reducing Peripheral Vision in Figure \ref{fig:possible_solutions_visual_attention}). 
    Dijana reported experiencing symptoms of visual vertigo and nausea as a result of this method.
    Similarly, Nicola expressed significant discomfort, stating: \textit{“With my hemianopsia, I only have half of my vision, and if you take away more of my vision, that frightens me. I know that when I walk through the world I am already missing three quarters of the world.”}

    \item Nicola expressed skepticism regarding the general usefulness of visual markers designed to capture attention. 
    She noted that in many cases, being unaware of certain elements in her environment was not problematic, and that being visually prompted about numerous items could become overwhelming:
    \begin{quote}
    \textit{“I’m blissfully unaware […] of the things in my scene. I’m not bothered by the fact that I don’t see everything, and all of a sudden seeing a lot of things can be stressful.”}
    \end{quote}
    Nevertheless, she acknowledged that such markers could be beneficial if the system were context-aware and able to selectively highlight only the most relevant objects.
    
    \item Dijana echoed similar concerns about overusing visual prompts. 
    She cautioned that excessive exposure to markers could lead to decreased effectiveness, as prompts might become visually assimilated into the scene: \textit{“Don’t keep the prompt for longer, because once it’s there in the scene it can become part of it.”}
    She emphasised the need to minimise visual clutter by limiting markers to only the most critical items. 
    Eye-tracking was proposed as a potential strategy to enable context-aware presentation of markers, but technical constraints prevented further exploration within this study.
    These reflections suggest that attention-guiding interventions must strike a careful balance between providing support and avoiding cognitive intrusion—especially for people with CVI. 
\end{itemize}


\begin{itemize}[label=\key, leftmargin=*, itemsep=2pt]
\item 
Co-designers felt that while attention-guiding features—like the wagon wheel technique or visual markers—were not effective as real-time assistive tools, they showed promise as training devices. 
These methods helped illustrate how to direct attention, but often caused discomfort or overload in everyday use. 
They also highlighted the need for systems that support learning without becoming burdensome during everyday tasks.
Co-designers found that smart glasses could be a tool to teach people how to guide their attention, using methods like the wagon wheel.
They also found visual markers helped guide attention, but also introduced discomfort and cognitive overload. 
Co-designers emphasised the need for personalised, context-aware systems that balance support with sensory limits—especially in cluttered or visually demanding environments.

\end{itemize}

\begin{figure*}[t]
    \centering
    \includegraphics[width=\textwidth]{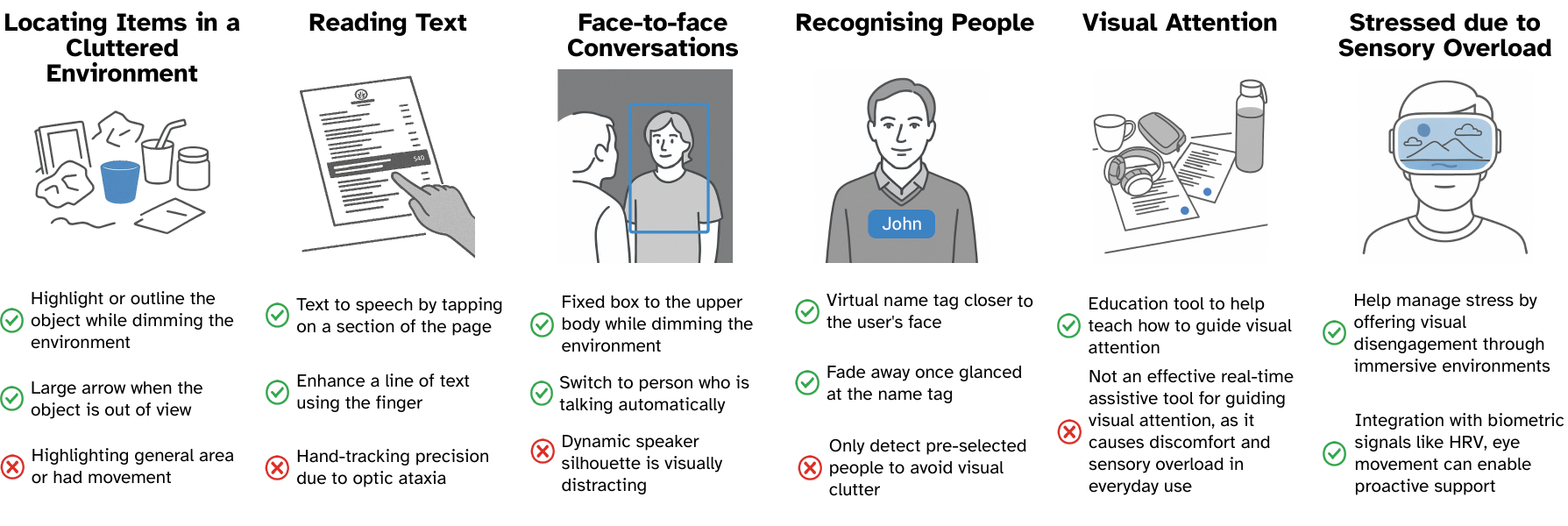}
    \caption{Summary of findings from the development workshops for the six identified difficulties.}
    \label{fig:findings_summary}
    \Description{Image showing summary of findings from the development workshops for the six identified difficulties, each includes a title, image, and two to three findings lablled as liked and disliked.
Locating Items in a Cluttered Environment
Image: a cluttered surface with paper, bottles, and containers. One cup is highlighted in blue.
liked: Highlight or outline the object while dimming the environment
liked: Large arrow when the object is out of view
disliked: Highlighting general area or head movement

Reading Text
image: a hand points to a section of a printed document. One line is bolded where the index finger is.
liked: Text to speech by tapping on a section of the page
liked: Enhance a line of text using the finger
disliked: Hand-tracking precision due to optic ataxia

Face-to-face Conversations
image: two people face each other; the person being spoken to has a blue box around their upper body and head.
liked: Fixed box to the upper body while dimming the environment
liked: Switch to person who is talking automatically
disliked: Dynamic speaker silhouette is visually distracting

Recognising People
image: a person faces forward with a virtual blue name tag labelled “John” floating under their chin.
liked: Virtual name tag closer to the user’s face
liked: Fade away once glanced at the name tag
disliked: Only detect pre-selected people to avoid visual clutter

Visual Attention
image: a cluttered tabletop with documents, cups, and a bottle. Blue dots indicate markers to guide visual attention.
liked: Education tool to help teach how to guide visual attention
disliked: Not an effective real-time assistive tool for guiding visual attention, as it causes discomfort and sensory overload in everyday use

Stressed due to Sensory Overload
image: a person wears a headset showing a calming virtual scene with mountains and clouds.
liked: Help manage stress by offering visual disengagement through immersive environments
liked: Integration with biometric signals like HRV, eye movement can enable proactive support
}
\end{figure*}

\subsubsection*{\textbf{Stressed due to sensory overload:}}
The goal of this design exploration was not to support specific daily tasks, but to help co-designers manage and mitigate sensory-induced stress—particularly to prevent what they described as a `CVI meltdown', a state of overwhelming sensory distress \cite{mcdowell2019personal, mcdowell2020cvi}.
A total of 16 design options were explored.

\begin{itemize}[label=\tick, leftmargin=*, itemsep=2pt]

\item Both co-designers expressed a preference for having the ability to completely block visual input from the external environment using the smart glasses. 
This was described by Nicola as \textit{“taking a calm break in a calm space,”} emphasising the need for sensory block in moments of overload.

\item Their preferred method for achieving this involved using the pre-built immersive virtual environments available on the device platform (see Figure \ref{fig:possible_solutions_stressed}). 
    Nicola commented:\textit{“It’s useful when you need to totally shut off from the world.”}
    For Dijana, these environments had a direct impact on emotional regulation: \textit{“
    It calms the mental and emotional states.”}

\item Both emphasised the potential of the device to serve a proactive role in detecting physiological signs of stress and intervening before sensory overload occurs.
    Nicola noted a correlation between her CVI meltdowns and a drop in heart rate variability (HRV), a known biomarker of stress \cite{rowe1998heart, immanuel2023heart, kim2018stress, kim2024reduced}. 
    Dijana similarly reported physical exhaustion due to visual fatigue. 
    These observations suggest that integrating biometric data from connected wearables, such as smartwatches, could enable real-time stress monitoring and early intervention.

\item The co-designers also responded positively to the Mindfulness app included on the device platform (Figure \ref{fig:possible_solutions_stressed}). 
    This application guides users through a breathing exercise. 
    Both reported feeling calmer after using the app. 
    However, further empirical validation is needed to assess its long-term effectiveness.

\item Since full visual block is only feasible when stationary, both co-designers preferred guided sensory reduction with adjustable modes—\textit{Minimal, Medium, and Extreme}—that could adapt to different contexts. For \textit{Minimal}, they preferred options like white noise or darkening the environment with ambient environment sound. \textit{Medium} involved fully darkening the environment or using the Mindfulness app. \textit{Extreme} combined immersive visual environments with noise cancellation. They expressed interest in customising each mode as they became more familiar with the glasses.

\end{itemize}

\begin{itemize}[label=\cross, leftmargin=*, itemsep=2pt]

\item An initial design goal was to reduce sensory overload by filtering out irrelevant visual information from the user’s environment. 
However, a major limitation of the device platform was the inability to directly modify the pass-through video feed from the onboard camera. 
Attempts to obscure elements in the real-world view using virtual overlays often resulted in additional visual clutter, undermining the goal of simplification.
This highlights a broader challenge in AR-based assistive design: techniques intended to simplify the visual field must avoid introducing new sources of clutter or distraction.
\end{itemize}


\begin{itemize}[label=\key, leftmargin=*, itemsep=2pt]
\item Both co-designers recognised that smart glasses help manage stress by offering visual disengagement through immersive environments and mindfulness tools. 
Future integration with biometric signals like HRV may enable proactive, personalised support to prevent CVI meltdowns and promote emotional well-being.
\end{itemize}


\subsection{Outlook and Optimism}
Both co-designers expressed strong optimism about smart glasses as a visual aid for people with CVI.
Nicola, initially skeptical, became enthusiastic by the study’s end:
\begin{quote}
\textit{“Initially I was skeptical about the usefulness of the glasses, but by the end, I want to buy one of these and cannot wait for it [smart glasses with the explored solutions] to be commercially available.”}
\end{quote}
This enthusiasm echoes broader findings, where OVI users reported improved independence and quality of life with smart glasses \cite{varshney2025evaluating}.
However, issues such as affordability and technical limitations remain barriers to broader adoption \cite{who_assistive_technology_2024}.


%% file: 6_discussion.tex
\section{Discussion}
This section discusses the findings from the study (summarised in Figure \ref{fig:findings_summary}) and explores their implications for smart glass research for people with CVI.
We aim to answer our primary research question: \textbf{How can smart glasses alleviate current challenges faced by people with CVI when understanding and interacting with their immediate environment?}  

\subsection{For which challenges are smart glasses useful?}

Our findings show that smart glasses can successfully support people with CVI to manage  challenges such as locating items in cluttered environments, reading text, recognising people, and facilitating face-to-face conversations. 
Co-designers highlighted a secondary benefit in mitigating stress caused by sensory overload. 
However, features aimed at directing visual attention were seen as less effective and potentially adding to cognitive load. 
Instead, these were valued more as training tools for learning how to guide visual attention.

For locating, many of the co-designers’ visual preferences—such as static highlights and darkened backgrounds for contrast—mirrored those reported by people with low vision \cite{zhao2016cuesee}
However, unlike people with low vision, our co-designers did not find auditory cues helpful for locating objects. 
This aligns with prior research on CVI, which highlights challenges in integrating spatial audio information \cite{philip2014identifying, lam2010cerebral, assets2024}.

For reading text, while prior work has explored smartphone-based text reading aids \cite{guo2016vizlens, tan2024exploration}, few studies have examined in-situ text augmentation using AR \cite{zhao2015foresee}. 
Our work extends this by embedding both visual and auditory support directly into the physical reading environment—a combination both co-designers preferred.

With respect to face-to-face communication, to the best of our knowledge, this is the first study exploring the use of smart glasses to support conversations by reducing visual clutter and background noise.

For recognising people, AT research  has primarily focused on providing auditory identification cues \cite{daescu2019deep, zhao2018face, facerecog1, facerecog2}. 
To our knowledge, there has been limited exploration of visual-only identification aids.
One notable exception is the work by Jain et al. \cite{jain2025social}, who proposed an augmented reality visual aid to support face recognition for people with prosopagnosia\footnote{Prosopagnosia, or face blindness, is a cognitive condition that impairs the ability to recognise familiar faces despite otherwise intact visual and cognitive abilities.}. 
However, their study primarily focused on the development of the underlying machine learning system and did not involve validation with end users affected by prosopagnosia.    

All techniques for guiding visual attention explored in our study involved manipulating virtual elements overlaid between the user’s eyes and the device platform’s pass-through video. 
This approach aligns with other studies in the field, which have used similar methods for guiding visual search \cite{lu2012subtle} and navigation \cite{koch2014natural}. 
However, prior research has investigated saliency modulation directly on the pass-through video to influence attention \cite{sutton2022look, sutton2024flicker, veas2011directing, mendez2010focus}. 
Due to device platform limitations discussed in Section \ref{sec:device platform}, we were unable to explore direct modification of the pass-through video. 
Future work should investigate the effectiveness of such approaches for supporting visual attention in people with CVI.

To manage stress caused by sensory overload for people with CVI, we believe our research is the first to specifically explore the use of immersive virtual environments in smart glasses. 
While virtual reality has been widely studied in the context of stress reduction and sensory regulation~\cite{soyka2016enhancing, velana2022advances, naef2023effects}, its targeted application for CVI populations remains under-explored. 
We believe this study represents an early step toward understanding how stress reduction interventions may indirectly enhance daily functioning for people with CVI.
This work aligns with growing interest in designing technologies that support emotional resilience for neurodiverse people ~\cite{stefanidi2024teenworlds, le2024human}.



\subsubsection*{\textbf{Does this generalise to other challenges?}}
While we investigated only six difficulties faced by people with CVI, we believe similar designs will be of benefit for many of the other previously identified challenges~\cite{assets2024}.  
During the iterative development process, we observed that designs intended to target one difficulty often offered benefits for others. 

For instance, the use of visual highlighting to locate items (\texttt{C2- Locating}) also proved useful for drawing attention to salient objects in busy visual scenes (\texttt{C5-Sensory Overload}).
Visual markers used to guide or redirect attention (\texttt{C5-Sensory Overload}) may enhance situational awareness (\texttt{C1-Unawareness}), as suggested by prior work \cite{lu2012subtle}. 
Similarly, edge-based highlighting techniques could also benefit users with difficulties in subject and background separation (\texttt{C7-Luminance and Contrast Sensitivity}). 
Likewise, the use of 3D directional arrows (\texttt{C2-Locating}) were discussed by our co-designers as potentially suited to navigation (\texttt{C6-Mobility}),
and  the solutions we developed to support recognising people (\texttt{C3-Recognising}) was also envisioned for locating people, e.g., asking the glasses to “Find Wendy,” with the device displaying a name tag when Wendy was identified—bridging \texttt{C3-Recognising} and \texttt{C2-Locating}.

The cross-challenge value of many designs may stem from their alignment with the broader cognitive-perceptual patterns of high-level visual difficulties.
While smart glasses have been widely explored for supporting lower-level visual difficulties such as visual field and visual acuity \cite{kasowski2023systematic}, there is little research into their role in addressing higher-level visual difficulties \cite{gamage2024broadening}.

Both co-designers in our study were diagnosed with simultanagnosia—a high-level visual difficulty characterised by an inability to perceive multiple objects simultaneously. 
This condition has been shown to impact all six of the selected difficulties explored in our study \cite{dalrymple2007seeing, baylis1994reading, friedman1995parietal, thomas2009reduced, rizzo2002psychoanatomical, karnath2005cortical}.
More recent findings also describe a dynamic constriction of visual attention under environmental complexity \cite{st2025emulation}, highlighting the evolving nature of how these conditions are understood.

Notably, both co-designers reported that the smart glasses helped alleviate some aspects of their simultanagnosia. 
This aligns with prior neurological work in simultanagnosia, that explored the use of visual properties (e.g., contrast, colour) to bias visual processing toward other visual pathways \cite{thomas2012enabling}. 
Our solutions—many of which rely on strong visual cues such as highlighting and overlays—may align with these visual pathways, potentially supporting or even modulating perception at a neurological level.
Dijana, for example, described the impact of the virtual box added during conversations as: \textit{"Something in my visual pathway falls into place because of this."}

While promising, these findings must be interpreted with caution.
Simultanagnosia is only one of many higher-level visual difficulties observed in people with CVI. 
Other conditions—such as akinetopsia (motion blindness)\footnote{An inability to perceive motion, resulting in the world appearing as static frames.}, palinopsia\footnote{Visual afterimages that persist even after the object has moved.}, and visual snow\footnote{A persistent visual disturbance resembling TV static or flickering dots.}—present fundamentally different perceptual challenges and often co-occur with complex neurological conditions such as cerebral palsy \cite{lueck2015vision, martin2016cerebral}.

Because both co-designers in this study shared a similar diagnosis, our design process and findings were naturally shaped around their specific perceptual experiences. 
As such, while our results suggest that certain visual strategies (e.g., strong colour contrast, spatial overlays) may benefit people with simultanagnosia, their applicability to other high-level visual impairments remains uncertain.


\subsubsection*{\textbf{Can smart glasses also serve as a rehabilitation tool for people with CVI?}}

Prior research highlights the role of neuroplasticity in reshaping visual processing through targeted interventions \cite{bennett2020neuroplasticity, delay2023interventions, mcdowell2023review, weden2022evidence, waddington2017review, ciman2013helpme}. For instance,  prior studies on perceptual learning show that targeted exposure can improve face recognition by people with prosopagnosia \cite{byrne2025rehabilitation}.
While neuroplasticity alone cannot overcome CVI—a lifelong condition rooted in neurological differences—it can support adaptation by strengthening compensatory pathways and strategies over time.

Both co-designers speculated that repeated use of smart glasses might contribute to such adaptation, helping users learn to manage their visual experiences more effectively. 
Dijana articulated this potential explicitly:
    \begin{quote}
    \textit{"I genuinely think with the right input, some people with CVI could be trained for their brain to go into how it should be."}
    \end{quote}
While these reflections are speculative, they align with emerging VR-based research for people with CVI \cite{manley2022assessing} that explores how immersive virtual environments can promote perceptual learning.
Importantly, this is not about reversing CVI, but about learning to live with it—developing tools and strategies that make it easier to interpret and interact with the world.

These reflections highlight the fluid boundary between assistive technology and rehabilitative intervention. 
For certain difficulties—like locating items in cluttered environments—smart glasses were perceived by our co-designers as potentially serving an assistive role, offering compensatory support for abilities unlikely to be restored. 
In contrast, for difficulties like guiding visual attention, the device was viewed as a potential rehabilitative aid, helping users develop attentional strategies over time.
The co-designers suggested that the  role of the device might change over time and vary by user context. 
Younger users, in particular, might prioritise long-term skill development and cognitive adaptation, while older users might focus on immediate, task-oriented support. 
Future assistive technologies should accommodate this fluidity—supporting both real-time assistance and long-term perceptual development, depending upon the user's goals. 

\subsection{What are the key considerations and challenges for smart glasses to be an effective assistive tool?}
While our study highlights the potential of smart glasses to address key challenges faced by people with CVI, we cannot fully answer our primary research question without first examining the design and implementation barriers that shape their real-world effectiveness. 
This section outlines the critical considerations that must be addressed for smart glasses to function as reliable and impactful assistive technologies.

\subsubsection*{\textbf{Hyper-Personalisation:}}
CVI is not a single condition but an umbrella term encompassing a wide range of brain-based visual processing difficulties \cite{Dutton2003Cognitive, roman2007cortical, lueck2019profiling}. 
People with CVI may present with varied impairments, including both lower and higher-level visual processing impairments \cite{philip2014identifying, lueck2019profiling}. 
In the following sections, we illustrate the need for hyper-personalisation informed design by considering user-specific visual conditions and context-aware adaptations.

\subsubsection*{User-Specific Visual Conditions:}
Even within our small co-designer group, solution preferences varied due to differences in visual impairments. 
For example, overlays for locating objects had to avoid the right visual field for Nicola due to her hemianopsia. 
Similarly, determining what constitutes a `salient' feature depends on individual attention profiles—for instance, a tendency to miss blue-coloured objects could warrant targeted visual markers only on them. 
The onset and developmental history of CVI also shaped interaction preferences. 
Dijana benefited from the face-to-face conversation aid, while Nicola had already developed robust compensatory strategies, reducing the perceived value of it. 
These observations underscore the limitations of one-size-fits-all approaches and highlight the need for highly adaptive systems \cite{aflatoony2022one, wey2004one, hodson2024understanding}.

\subsubsection*{Context-Aware Adaptation:}
Smart glasses are uniquely positioned to interpret contextual cues, given their alignment with the user’s field of view \cite{Kim2021Applications}.
Throughout the study, context-awareness emerged as a critical requirement for adapting augmentations effectively. 
For example, highlighting required dynamic adjustments to environmental lighting—brightening visuals in dim settings while avoiding over-illumination in bright ones. 
Similarly, text-to-speech speed needed to adapt to background noise levels, slowing down in distracting environments and speeding up in quieter contexts.

Both co-designers expressed interest in integrating physiological indicators—such as heart rate variability \cite{rowe1998heart}, eye movements\cite{jacob1990you}, or pupil dilation \cite{pomplun2019pupil}—to inform real-time adjustments. 
These bio-signals are increasingly recognised as reliable proxies for cognitive load and emotional state \cite{schultz2023biosignals, schmidt2018wearable}, offering a promising pathway towards truly adaptive and responsive assistive experiences.
This points to the value of designing adaptive systems that use real-time signals—not just from the environment but also from the user—to modulate assistive strategies dynamically.

\subsubsection*{\textbf{Usability Consideration:}}

While general accessibility principles remain applicable \cite{blvdesignconsiderations}, several CVI specific usability concerns emerged.

\subsubsection*{Presentation modalities:}
Generally co-designers preferred environmental information to be presented visually. 
This aligns with prior literature on CVI assistive needs \cite{gamage2024broadening, assets2024}. 
Smart glasses enable this through visual overlays such as highlights and name tags, embedding directly into the user’s environment.
This is in contrast to devices like smartphones, where augmentations can only be viewed indirectly on a 2D screen.
As Nicola reflected:
\begin{quote}
\textit{"These kinds of visual overlays—like the highlight around the object—really help. When it outlines the exact shape, it makes it easier to judge size and position, which is what I need to grab things properly. It shows where the object is, so it feels natural." }
\end{quote}

In contrast, however, for the two tasks involving language--reading and face-to-face conversation--there was a preference for the conversation and text to be presented both visually and aurally.
This is interesting because prior research has found no effect of mode of input (listening, reading, or listening and reading simultaneously) on verbal comprehension by adults who do not have CVI~\cite{rogowsky2016does}. 
More research is required to investigate if, in fact, dual presentation is beneficial and whether it reduces cognitive load.

\subsubsection*{Interaction Modalities:}
For control input, co-designers favoured voice interaction.
Using SiriKit, we enabled simple voice commands to trigger specific assistive functions (\textit{e.g., `Hey Siri, find my cup.'}). 
While these were limited to predefined actions, this approach proved both intuitive and accessible.
This preference is in line with broader accessibility literature on natural language interaction for people with vision impairments \cite{voicecontrolled1, speech_input_blind}.

\subsubsection*{Seamless Integration:}
Beyond the ability to issue commands, co-designers expressed a need for the system to anticipate and activate appropriate tools automatically. 
Rather than requiring constant manual input, they preferred tools that could activate and deactivate based on the situation—reducing the need for ongoing interaction.
For instance, we implemented a mechanism that automatically stopped object highlighting once the user’s hand touched the target. 
This served both to confirm successful interaction and to minimise visual clutter.
In the context of CVI, where cognitive and sensory load are already high, these subtle system adaptations play a critical role in enhancing usability and promoting sustained use.

\subsubsection*{The Role of Colour:}
Colour preferences were highly personal but showed consistent cognitive implications. 
Racey et al. \cite{racey2019processing} found that preferred colours can bias attention, memory encoding, and aesthetic perception.
In our study, this was reflected in Dijana’s preference for pastel orange and Nicola’s for cyan—both co-designers reported that using their preferred colours in visual augmentations enhanced clarity and overall comfort.

At the same time, colours must remain legible in diverse visual environments. 
Following accessibility best practices, colour choices should adapt dynamically based on background luminance and contrast requirements to maintain visibility and reduce perceptual strain.

\subsubsection*{Managing Cognitive Load:}
An important consideration was the need to minimise visual clutter.
Despite a general preference for visual modalities, moving, flashing, or overly dense visualisations were poorly tolerated (refer to the preferred and disliked examples of highlighting in Figure \ref{fig:possible_solutions_highlighting}).
Co-designers reported that they either ignored such elements, or else they actively contributed to sensory overload.
This aligns with prior work showing that poorly designed AR augmentations can lead to confusion, fatigue, and frustration—even among people with typical vision \cite{duan2022confusing, koppel2021context, jankowski2019gradual}.
This underscores the need for minimal, non-intrusive, and spatially relevant augmentations—co-designed to align with personal cognitive thresholds \cite{codesign1}.

\subsubsection*{\textbf{Current Technological Limitations:}}
Like many emerging technologies, smart glasses face practical constraints, including performance in environmental variability, hardware limitations, and battery life \cite{Xu2018Sensor, Kim2021Applications}. These challenges are compounded by limitations in machine learning—such as reliability issues \cite{ml_accuracy} and accuracy constraints on low-power devices \cite{assets2024}. 
Nevertheless, the technology is evolving rapidly, with many of these issues steadily being addressed.

%% file: 7_limitations.tex
\section{Limitations and Future Work}
This study marks an important first step—an exploratory investigation into how smart glasses can support people with CVI. However, much work remains to be done.

A key limitation of this study is the small sample size (n=2). 
While this allowed for deep, individualised engagement and the development of highly personalised solutions, it limits the generalisability of our findings. 
The design decisions and outcomes presented here may not translate directly to the broader CVI population with other neurological conditions.
Future research should involve a larger, more diverse CVI cohort and integrate quantitative task-based performance measures with qualitative insights.

As identified earlier, there are several promising directions for future work, including exploring additional challenges, expanding the range of possible solutions, and, in particular, understanding how smart glasses can be leveraged to help reduce sensory overload.

In addition, while we began to explore how users might engage with the smart glasses as a unified system—with features like automatic switching and Siri—this area remains underdeveloped. 
More research is required to understand how users manage multiple concurrent augmentations, and how the system can intelligently adapt to shifting contexts and user states.

%% file: 8_conclusion.tex
\section{Conclusion}
This paper explored how smart glasses can support people with CVI understand and interact with their environment.  Using a co-design process with two adults with CVI, we found that smart glasses can  support people with CVI in addressing key challenges—such as locating objects, recognising people, reading text, engaging in conversation and managing sensory stress—through spatially embedded visual augmentations like highlights, name tags, and virtual boxes. We also found that smart glasses hold promise as potential rehabilitation aids to teach strategies for controlling visual attention. 

We identified a number of design considerations for effective deployment: hyper-personalisation, context-aware adaptation, and strategies to minimise cognitive load. We found that while co-designers generally preferred visual output for presenting environmental information, for tasks involving language they preferred dual presentation as text and audio.


We believe our research highlights a timely and impactful area for further research. 
With the rapid advancement of wearable XR, there is significant opportunity to develop assistive technologies that dramatically improve the quality of life for people with CVI, a condition that is increasingly recognised as a major cause of vision impairment. 



%% file: 9_appendix.tex
\section{Analysis from the Scoping and Ideation Workshops}
\begin{figure}[H]
    \centering
    \includegraphics[width=0.75\textwidth]{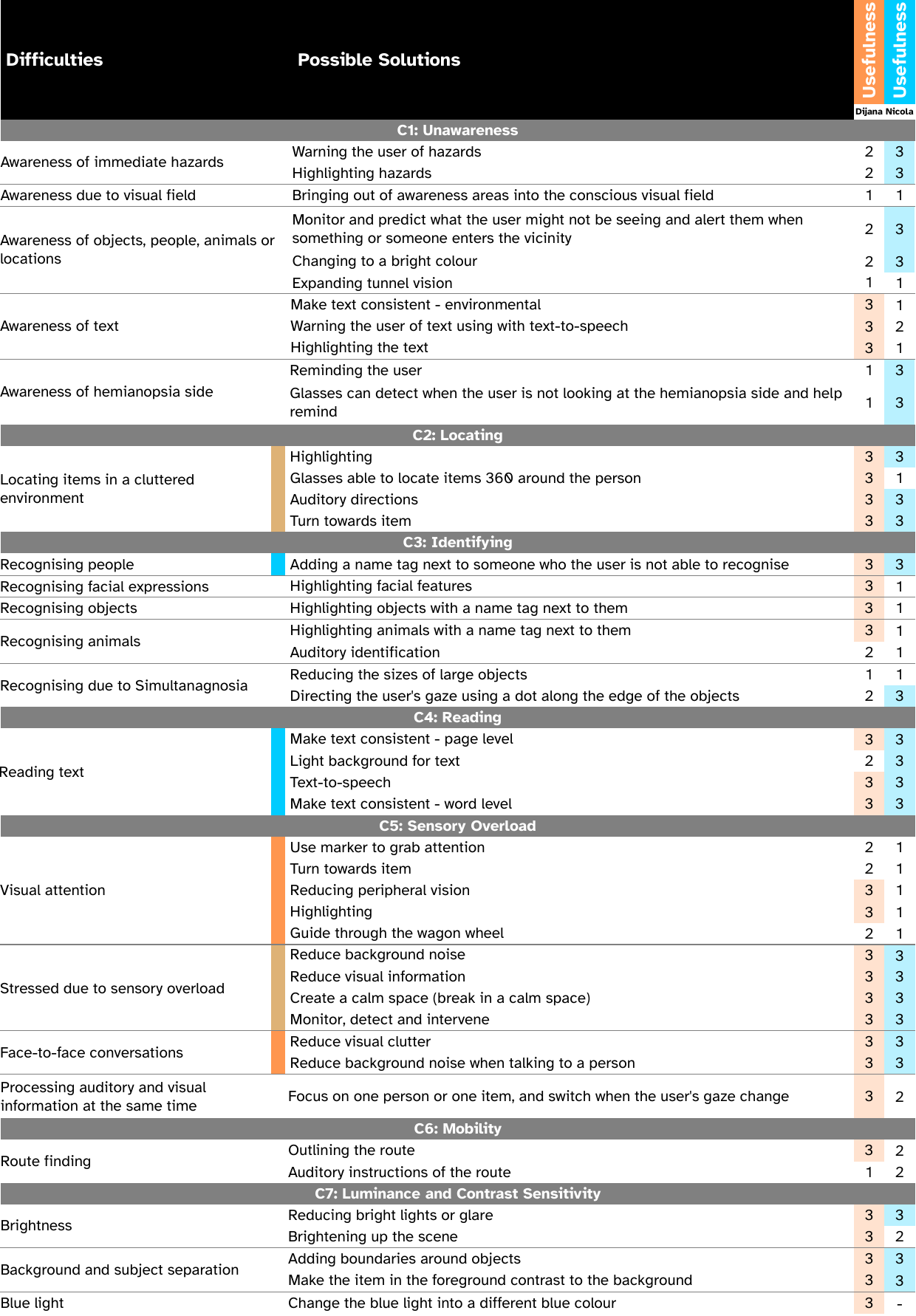}
    \caption{Analysis from the Scoping and Ideation workshops showing identified difficulties, possible smart-glass solutions, and usefulness scores provided by each co-designer. 
    The brown bars represent difficulties selected by both co-designers, cyan indicates selections by Nicola, and pastel orange indicates selections by Dijana. 
    Usefulness scores (1 = not useful, 2 = somewhat useful, 3 = very useful) are shown for each solution, reflecting individual co-designer feedback.}
    \label{fig:difficulties_possible_solutions}
    \Description{Refer to supplementary materials Figure10.csv for the table text.}
\end{figure}

\section{Overview of Development Workshops}
\begin{figure}[!h]
    \centering
    \includegraphics[width=\textwidth]{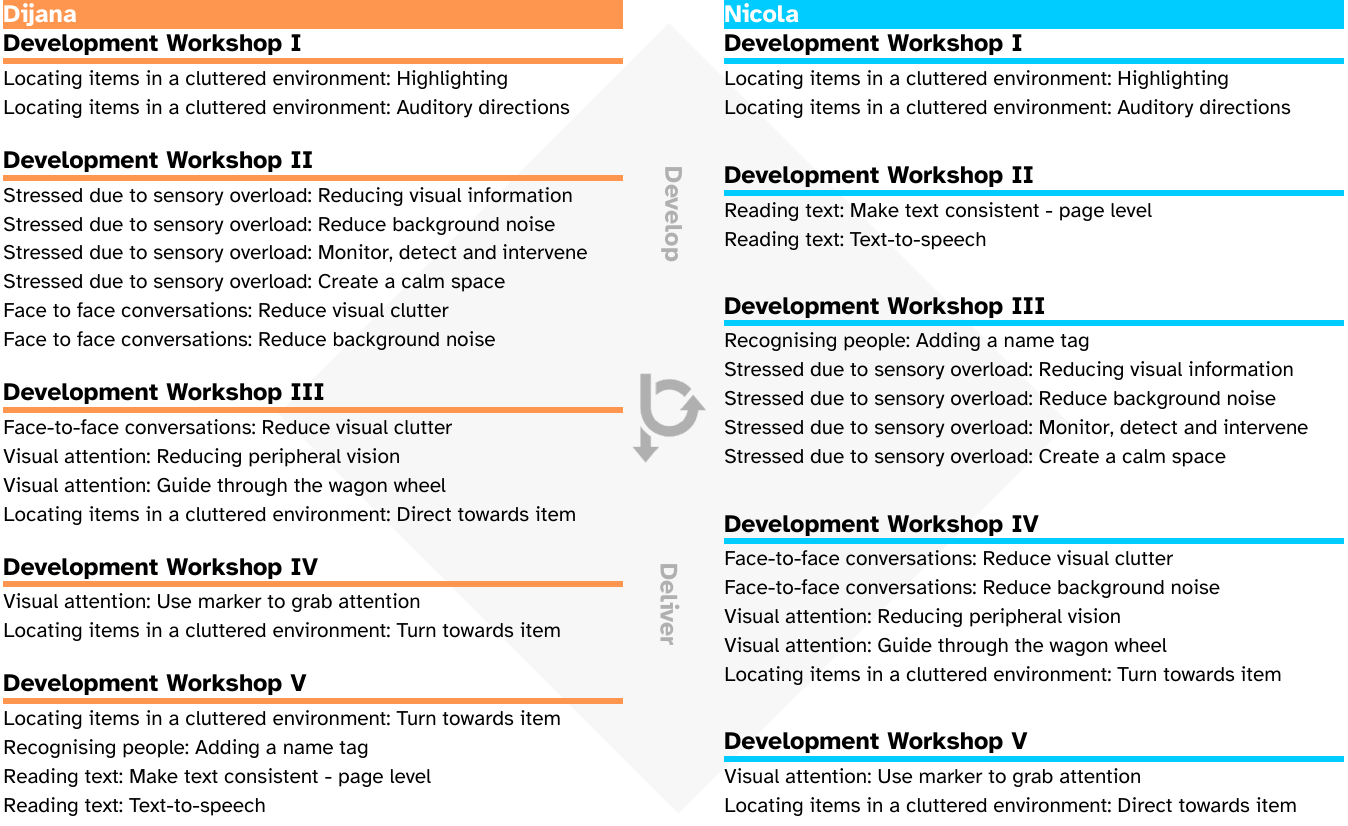}
    \caption{Overview of possible solutions discussed during each development workshop with the two co-designers. The figure highlights the primary focus areas for each session, although discussions often revisited and refined ideas from earlier workshops. Colour coding distinguishes between Dijana’s (pastel orange) and Nicola’s (cyan) sessions.}
    \label{fig:methodology_overview}
    \Description{Lists of development workshops for Dijana and Nicola. A diamond is shown in the background, progressing from develop to deliver.
        Dijana
        Development Workshop I
        Locating items in a cluttered environment: Highlighting
        Locating items in a cluttered environment: Auditory directions
        Development Workshop II
        Stressed due to sensory overload: Reducing visual information
        Stressed due to sensory overload: Reduce background noise
        Stressed due to sensory overload: Monitor, direct and intervene
        Stressed due to sensory overload: Create a calm space
        Face to face conversations: Reduce visual clutter
        Face to face conversations: Reduce background noise
        Development Workshop III
        Face to face conversations: Reduce visual clutter
        Visual attention: Reducing peripheral vision
        Locating items in a cluttered environment: Direct towards item
        Development Workshop IV
        Visual attention: Use marker to grab attention
        Locating items in a cluttered environment: Turn towards item
        Development Workshop V
        Locating items in a cluttered environment: Turn towards item
        Recognising people: Adding a name tag
        Reading text: Make text consistent - page level
        Reading text: Text-to-speech
        Nicola
        Development Workshop I
        Locating items in a cluttered environment: Highlighting
        Locating items in a cluttered environment: Auditory directions
        Development Workshop II
        Reading text: Make text consistent - page level
        Reading text: Text-to-speech
        Development Workshop III
        Recognising people: Adding a name tag
        Stressed due to sensory overload: Reducing visual information
        Stressed due to sensory overload: Reduce background noise
        Stressed due to sensory overload: Monitor, direct and intervene
        Stressed due to sensory overload: Create a calm space
        Development Workshop IV
        Face to face conversations: Reduce visual clutter
        Face to face conversations: Reduce background noise
        Visual attention: Reducing peripheral vision
        Visual attention: Guide through the wagon wheel
        Locating items in a cluttered environment: Turn towards item
        Development Workshop V
        Visual attention: Use marker to grab attention
        Locating items in a cluttered environment: Direct towards item
        }
\end{figure}